\documentclass[10pt,amsmath,amsfonts,amssymb,aps,twocolumn,floatfix,prl]{revtex4-1}
\pdfoutput=1
\usepackage[utf8]{inputenc}
\usepackage[T1]{fontenc}
\usepackage[english]{babel}
\usepackage{csquotes}

\usepackage{graphicx}

\usepackage{upgreek}
\usepackage[cspex,bbgreekl]{mathbbol}

\newcommand{\ket}[1]{\left| #1 \right>} % for Dirac bras
\newcommand{\bra}[1]{\left< #1 \right|} % for Dirac kets

\def\ifnonemptyparenthesis#1{%
  \if\relax\detokenize{#1}\relax%
  \else%
    (#1)%
  \fi%
}
\newcommand{\sub}[1]{_{\mathrm{#1}}}
\newcommand{\fig}[2][]{Fig.\,\ref{#2}\ifnonemptyparenthesis{#1}}
\newcommand{\eq}[1]{Eq.\,\ref{#1}}
\newcommand {\ii} {\mathrm{i}}

\newcommand{\invisiblesection}[1]{%
  \phantomsection%
  \stepcounter{section}%
  \addcontentsline{toc}{section}{\protect\numberline{\thesection}#1}%
  }
		
\usepackage[colorlinks=true,citecolor=blue,linkcolor=black,urlcolor=blue]{hyperref}

\begin{document}
\title{Spin pumping and measurement of spin currents in optical superlattices}
\author{C.~Schweizer$^{1,2}$, M.~Lohse$^{1,2}$, R.~Citro$^{3,4}$, I. Bloch$^{1,2}$}
\affiliation{$^{1}$\,Fakult\"at f\"ur Physik, Ludwig-Maximilians-Universit\"at, Schellingstrasse 4, D-80799 M\"unchen, Germany\\
$^{2}$\,Max-Planck-Institut f\"ur Quantenoptik, Hans-Kopfermann-Strasse 1, D-85748 Garching, Germany\\
$^{3}$\,Dipartimento di Fisica ``E.~R.~Caianiello'', Università degli Studi di Salerno,
Via Giovanni Paolo II 132, I-84084 Fisciano (Salerno), Italy\\
$^{4}$\,SPIN-CNR Salerno, Via Giovanni Paolo II 132, I-84084 Fisciano (Salerno), Italy}

\begin{abstract}
We report on the experimental implementation of a spin pump with ultracold bosonic atoms in an optical superlattice.
In the limit of isolated double wells it represents a 1D dynamical version of the quantum spin Hall effect.
Starting from an antiferromagnetically ordered spin chain, we periodically vary the underlying spin-dependent Hamiltonian and observe a spin current without charge transport.
We demonstrate a novel detection method to measure spin currents in optical lattices via superexchange oscillations emerging after a projection onto static double wells.
Furthermore, we directly verify spin transport through in-situ measurements of the spins' center of mass displacement.
\end{abstract}

\date{\today}

%Ultracold gases, trapped gases. Phases: geometric; dynamic or topological.
%\pacs{67.85.-d, 03.65.Vf}

\maketitle

\invisiblesection{Introduction}
Exposing materials to strong magnetic fields has led to remarkable discoveries, most prominently the pioneering observation of the integer and fractional quantum Hall effect~\cite{Klitzing:1980, Tsui:1983}.
These quantum phenomena surprise due to their robustness and independence of material properties, arising from their topological nature~\cite{Thouless:1982}.
In this context, Thouless recognized that 1D dynamical systems can share the same topological character as the 2D integer quantum Hall (IQH) effect~\cite{Thouless:1983,Niu:1984}.
Such topological charge pumps exhibit a quantized transport per pump cycle in a gapped filled band of an adiabatically and periodically evolving potential.
More recently, a fundamentally different quantum state was observed~\cite{Koenig:2007}, the topological insulator (TI) \cite{Kane:2005, Bernevig:2006}, which preserves in addition to charge time-reversal symmetry. 
In 2D systems with spin conservation, it exhibits the quantum spin Hall (QSH) effect, characterized by a quantized spin but vanishing charge conductance.
Analogous to the Thouless pump, a dynamical version of a TI can be designed~--~a quantum spin pump~\cite{Shindou:2005, Fu:2006,Zhou:2014}.

Spin pumps could serve as spin current sources, e.g.\ for spintronic applications~\cite{Prinz:1998}.
Such spin current generators have been proposed based on the spin Hall effect~\cite{Murakami:2003,Kato:2004}, the cyclic variation of two system parameters in interacting quantum wires~\cite{Sharma:2001,Citro:2003}, and topological insulators~\cite{Citro:2011}. 
However, spin pumps have been realized only in few experiments, e.g.\ in quantum dot structures~\cite{Watson:2003} and by parametrically excited exchange magnons~\cite{Sandweg:2011}.
Here, we demonstrate the first implementation of a spin pump with ultracold bosonic atoms in optical superlattices and present a direct measurement of the arising spin current.

In 1983, Thouless investigated particle transport in two superimposed 1D periodic potentials adiabatically moved relative to each other .
This sliding motion periodically varies the combined potential and thereby the underlying single-particle Hamiltonian, which can be parametrized by the cyclic pump parameter~$\phi$.
During the pumping, a particle acquires an anomalous velocity proportional to the Berry curvature defined on the closed surface spanned by~$\phi$ and the quasi-momentum~$k$.
The resulting displacement after one pump cycle only depends on the geometric properties of the pump cycle and is not quantized unless all quasi-momenta of a band are occupied equally.
In this case, the pump is topological with a displacement proportional to the Chern number, the integral of the Berry curvature over the entire surface.
Thus, transport is quantized and robust against perturbations~\cite{Thouless:1983,Niu:1984}.
Recently, such geometric and quantized, topological pumps have been realized with ultracold bosonic~\cite{Lohse:2016, Lu:2016} and fermionic atoms~\cite{Nakajima:2016}.

In analogy to the Thouless pump, a spin pump can be thought of as a dynamical version of a QSH system  \cite{Shindou:2005}, which is characterized by a bulk excitation gap and gapless edge excitations. In general, the electron spin is not conserved, e.g.\ in the presence of spin-orbit coupling, and therefore unconventional topological invariants, like the $Z_2$ index~\cite{Kane:2005}, are needed for classification. 
Non-interacting QSH systems with spin conservation can be interpreted as two independent IQH systems.
Therefore, a quantum spin pump can be composed of two independent pumps, where the up and down spins have inverted Berry curvature and are therefore transported in opposite directions.

\invisiblesection{Implementation of a spin pump}
A quantum spin pump can be implemented with ultracold atoms in two hyperfine states in a spin-dependent dynamically controlled optical superlattice, which can be formed by superimposing two lattices with periods~$d_s$ and $d_\text{l}=2d_\text{s}$.
In the tight binding limit, a spin in this superlattice is described by the Rice-Mele model~\cite{Rice:1982}, which comprises staggered on-site energies~$\pm\Delta/2$ between neighboring sites and alternating tunnel couplings~$\frac{1}{2}\left(J\pm \delta J\right)$ with the dimerization parameter~$\delta J$.
Pumping can be induced by an adiabatic modulation of the potential and corresponds to a loop in parameter space ($\delta J$, $\Delta$) around the degeneracy point ($\delta J=0$, $\Delta=0$).
If the two spin components do not interact with each other, their pumping motion is independent and a spin pump can be realized by a spin-dependent deformation of the potential, so that time-reversal symmetry is retained and their Berry curvature is reversed. 
The associated spin transport is quantized only for equal occupation of all quasi-momenta, which can be realized with non-interacting fermions by placing the Fermi energy in the band gap and bosons by localizing each spin component to a Mott insulator with negligible inter-spin interaction.

In addition, the two spin components can be coupled by introducing on-site interactions~$U$ between the atoms.
For hardcore interactions and unit filling, the bare tunneling is suppressed and the system can be described by a 1D spin chain
\begin{equation}
	\begin{aligned}
		\hat{\mathcal{H}} =& -\frac{1}{4}\sum_{m}\bigl(J_\text{ex}+(-1)^{m}\delta J_\text{ex}\bigr)\left(\hat{S}^+_{m}\hat{S}^-_{m+1}+\text{h.c.}\right) \\
		&+ \frac{\Delta}{2}\sum_{m} (-1)^{m} \hat{S}^z_m
		\label{eq:spinmodel}
		\end{aligned}
\end{equation}
with spin-dependent tilt~$\Delta$ and alternating exchange coupling $\frac{1}{2}\left(J_\text{ex}\pm \delta J_\text{ex}\right) \simeq \left(J\pm \delta J\right)^2/U$.
For large tilts~$\Delta\gg\frac{1}{2}\left(J_\text{ex} + \delta J_\text{ex}\right)$ the many-body ground state are locked spins in an antiferromagnetic order, while for strong exchange coupling~$\frac{1}{2}\left(J_\text{ex} + \delta J_\text{ex}\right)\gg\Delta$ dimerized entangled pairs are favored.
Implementing this Hamiltonian requires a dynamically controllable spin-dependent superlattice~\cite{Lee:2007,Dai:2015}.
In the limit of isolated double wells $\delta J_\text{ex}\approx J_\text{ex}$, applying a global gradient to a spin-independent superlattice can locally reproduce the staggered tilts (\fig[a]{fig1}).
In this situation, a variation of the parameters ($\delta J$, $\Delta$) during the spin pump cycle corresponds to a modulation of ($J_\text{ex}$, $\Delta$) in the interacting 1D spin chain.
This cycle needs to be performed adiabatically with respect to the intra double well exchange coupling $\frac{1}{2}(J_\text{ex}+\delta J_\text{ex})$.
The pump cycle encircles the degeneracy point ($\delta J_\text{ex}=0$, $\Delta=0$) as illustrated in \fig[a]{fig1}.
Despite the antiferromagnetically ordered state being an excited many-body state in the globally tilted system, the pump can still be described by a topological invariant as the pump of \eq{eq:spinmodel} with local tilts, if the pumping  is fast compared to $\frac{1}{2}(J_\text{ex}-\delta J_\text{ex})$, which determines the gaps of additional level crossings in the spectrum~\cite{Supplementary}. 
This can be readily achieved in the experiment by choosing lattice depths, for which inter-double well coupling is quenched.

\begin{figure}[b]%
\includegraphics[width=\columnwidth]{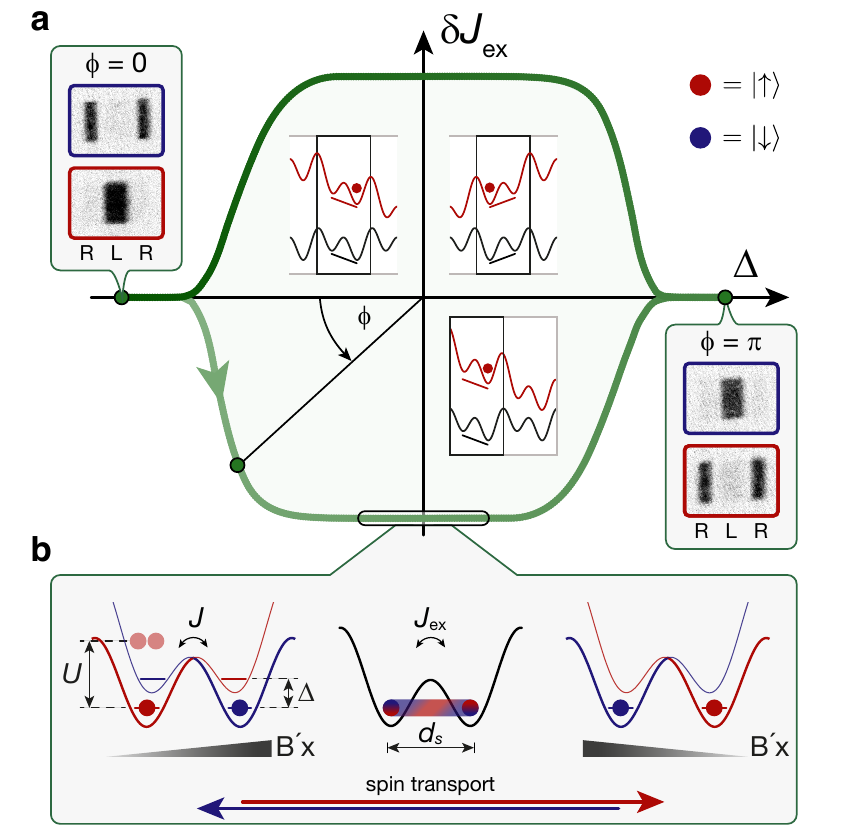}%
\caption{Spin pump cycle.
(a)~Spin pump cycle in parameter space (green) of spin-dependent tilt~$\Delta$ and exchange coupling dimerization~$\delta J_\text{ex}$. 
The path can be parametrized by the angle~$\phi$, the pump parameter. 
The insets in the quadrants show the local mapping of globally tilted double wells to the corresponding local superlattice tilts with the black rectangles indicating the decoupled double wells.  
Between $\phi=0$ and $\pi$, $\left|\uparrow\right>$ and $\left|\downarrow\right>$ spins exchange their position, which can be observed by site-resolved band mapping images detecting the spin occupation on the left (L) and right (R) sites, respectively. 
(b)~Evolution of the two-particle ground state in a double well around $\Delta=0$ with tunnel coupling~$\frac{1}{2}\left(J+\delta J\right)$, on-site interaction energy~$U$, and spin-dependent tilt~$\Delta$ as well as the exchange coupling $J_{ex}\simeq \left(J+\delta J\right)^2/U$ and the lattice constant~$d_s$.%
}%
\label{fig1}%
\end{figure}
\begin{figure*}%
\includegraphics[width=\textwidth]{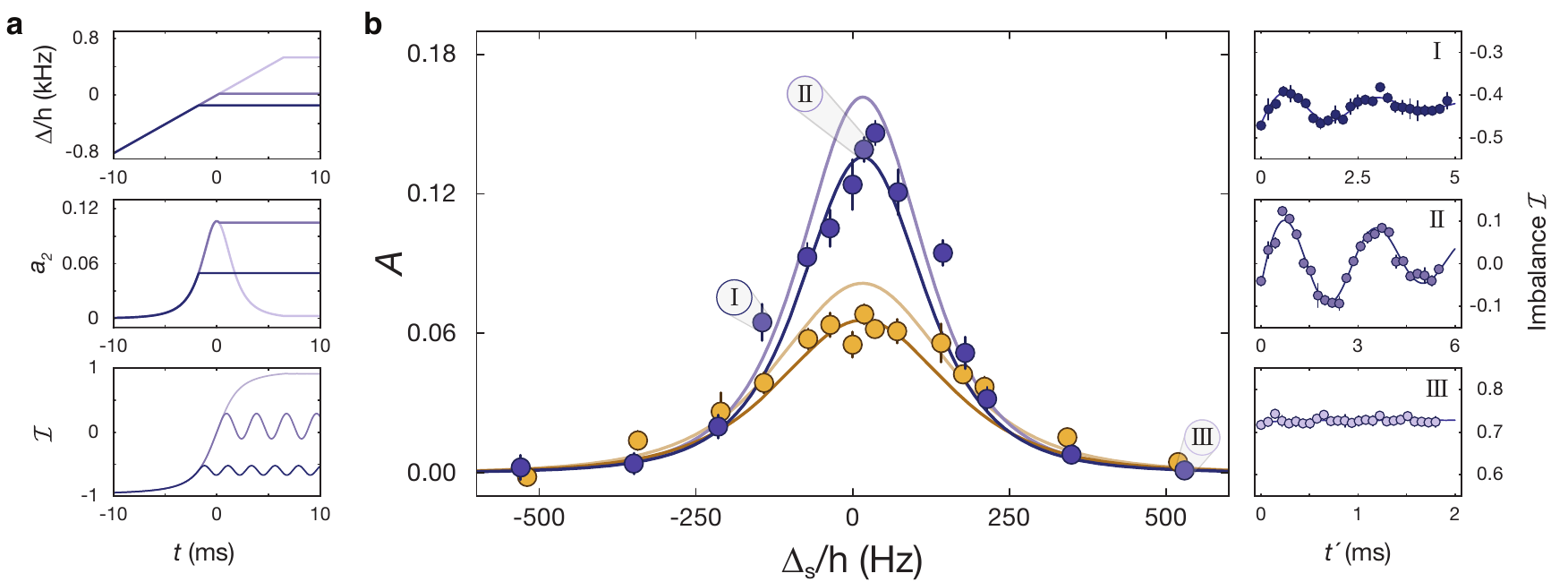}%
\caption{%
Spin current measurement.
(a)~Illustration of the measurement scheme. 
$\Delta$ is ramped with a rate of~$\dot{\Delta}=82(2)\,\text{kHz/s}$ at $\Delta=0$ and the ramp is stopped abruptly at different points in the cycle~$\Delta_\text{s}$ (upper graph). 
During this ramp, the two-particle wavefunction initially in the ground state has a small admixture~$a_2$ of the first excited state around $\Delta=0$ (middle graph).
After the stop of the ramp, this admixture leads to spin imbalance oscillations with an amplitude proportional to $a_2$ and thus to the instantaneous spin current at $t_\text{s}$.
The lower graph shows a numerical simulation of the spin imbalance time traces assuming perfect adiabaticity.
(b)~Spin imbalance oscillation amplitude~$A$ at different points in the pump cycle for $J_\text{ex}/h=342(2)\, \text{Hz}$ (blue) and $J_\text{ex}/h=467(3)\,\text{Hz}$ (orange).
Each point is the amplitude obtained by fitting \eq{eq:fitfunction} to the spin imbalance that was measured as a function of the holdtime~$t^\prime$; error bars are the fit uncertainty.
Three traces are shown on the right hand side for $\Delta_\text{s}/h=-144(7)\,\text{Hz}$ (I, dark blue), $\Delta_\text{s}/h=18(5)\text{Hz}$ (II, blue), and $\Delta_\text{s}/h=530(30)\,\text{Hz}$ (III, light blue) corresponding to the illustrations in (a).
Each trace consists of 26 points, which were averaged five times.
The light solid lines in the main plot show the numerical calculation for the oscillation amplitude taking into account the reduced detection efficiency due to a residual exponential decay of~$\Delta$.  
The dark solid lines include additionally a finite ground state occupation of~$97(1)\,\%$ and a pump efficiency of~$89(1)\,\%$, which were measured separately by band mapping.
\label{fig2}}%
\end{figure*}

\invisiblesection{Experimental realization}
The experimental setup consists of a 3D optical lattice with a superlattice along the $x$-axis and deep transverse lattices along $y$ and $z$ to create an array of decoupled 1D systems. 
Each system is initially occupied by an antiferromagnetically ordered spin chain of up $\left|\uparrow\right>=\left|F=1,m_{F}=-1\right>$ and down $\left|\downarrow\right>=\left|F=1,m_{F}=+1\right>$ $^{87}$Rb atoms~\cite{Widera:2005}, localized on individual sites with~$J_\text{ex}\approx\delta J_\text{ex}\approx 0$ \cite{Supplementary}.
To start the pump cycle, every second barrier is decreased to transfer two neighboring spins to the ground state of a double well. 
Since a large magnetic gradient $\Delta\gg J_\text{ex}$ is present along $x$, the up (down) spin stays localized on the left (right) side, shortly denoted as $\left|\uparrow,\downarrow\right>$.
The gradient is then reversed adiabatically compared to $J_\text{ex}$.
Thereby the wavefunction follows the instantaneous eigenstate and a spin current occurs as the two spins exchange their positions via the delocalized triplet state $\frac{1}{\sqrt{2}}\bigl(\left|\uparrow,\downarrow\right>+\left|\downarrow,\uparrow\right>\bigr)$ at $\Delta=0$ (\fig[b]{fig1}).
At the end of the first half cycle, individual sites are decoupled by increasing the short lattice depth.
Subsequently, $\delta J_\text{ex}$ is inverted by flipping the dimerization and also the magnetic gradient to maintain the correct local tilts (insets \fig[a]{fig1}).
This corresponds to a projection on double wells shifted by one lattice site.
The state of the system remains unchanged during this sudden switch and the spins in the new shifted double well are in the ground state. 
After a full cycle, the two spin components have each moved by $2\,d_s$ in opposite directions; therefore the total particle current vanishes as the contributions from the two spin components cancel each other exactly.
Thus, pumping leads to a spin transport without inducing a particle current.

\invisiblesection{Spin currents}
The spin current~$\mathfrak{j}$ between the left (L) and the right (R) site of a double well is related to the change in the expectation value of the spin imbalance $\mathcal{I}=\frac{1}{2}(n_{\text{L}\downarrow}-n_{\text{L}\uparrow}-n_{\text{R}\downarrow}+n_{\text{R}\uparrow})$ given by the integral form of the continuity equation $2\mathfrak{j}=\partial_{t} \mathcal{I}$, with $n_{i\sigma}$ the occupation of spin~$\sigma$ on site~$i$.
To understand how this spin current arises and how it can be detected, it is useful to examine the evolution of the eigenstates during the adiabatic change of~$\Delta(t)$.
Two spins initially at time~$t_\text{i}$ in the eigenstate $\left|n_{t_\text{i}}\right>$ of the two-particle double well Hamiltonian~$\hat{H}_\text{DW}(t_\text{i})$ \cite{Supplementary} follow the instantaneous eigenstate $\left|n_t\right>$, but acquire -- even for a perfect adiabatic evolution -- a small imaginary contribution~$\ii\,a_m(t)$ from other eigenstates~$\left|m_t\right>$. 
This admixture occurs only temporarily during the ramp and induces an anomalous spin velocity (\fig[a]{fig2}).
The coefficients $a_m$ can be calculated in first-order perturbation theory $a_{m}(t)=-\dot{\Delta}\frac{\left<m_t\right|\hbar\partial_\Delta\left|n_t\right>}{E_n(t)-E_m(t)}$ with~$\dot{\Delta}=\partial_t \Delta(t)$ the ramp speed and $E_l(t)$ the eigenenergy of~$\left|l_t\right>$~\cite{Xiao:2010}.
When starting from the ground state~$\left|1_t\right>$, the wavefunction is well approximated by only considering contributions from the first excited state~$\left|\psi_t\right>\approx \left|1_t\right>+\ii\,a_2(t)\left|2_t\right>$.
The $a_m$-coefficients of higher lying states are strongly suppressed as the corresponding wavefunctions depend weakly on $\Delta$ and $E_m-E_1 \gg J_\text{ex}$.
The wavefunction~$\left|\psi_t\right>$ can be probed by a sudden stop of the pump cycle at time~$t_\text{s}$ by projecting it onto~$\hat{H}_\text{DW}(t_s)$.
During the subsequent time evolution, the two states $\left|1_t\right>$ and $\left|2_t\right>$ acquire a relative phase leading to oscillations of the spin imbalance $\mathcal{I}(t)=\mathcal{I}_\text{s}+A \sin[(E_2(t_\text{s})-E_1(t_\text{s}))/\hbar\cdot(t-t_\text{s})]$ with~$\mathcal{I}_\text{s}$ the imbalance at time $t_\text{s}$.
The oscillation amplitude $A=-2\,a_2(t_\text{s})\left<1_{t_\text{s}}\right|\hat{\mathcal{I}}\left|2_{t_\text{s}}\right>$ is proportional to the admixture of the second eigenstate and can be related to the spin current~$\mathfrak{j}(t_\text{s})$ through the continuity equation~\cite{Supplementary}
\begin{equation}
\mathfrak{j}(t_\text{s})=A\,\frac{E_2(t_\text{s})-E_1(t_\text{s})}{2\hbar}.
\label{eq:CurrentToOscillationAmplitude}
\end{equation}
Experimentally, the gradient ramp was abruptly stopped at~$\Delta_\text{s}=\Delta(t_\text{s})$ and a time trace of the resulting double well superexchange oscillation~\cite{Trotzky:2007} was recorded by a simultaneous measurement of $n_{\text{L}\uparrow}$, $n_{\text{L}\downarrow}$, $n_{\text{R}\uparrow}$ and $n_{\text{R}\downarrow}$ with Stern-Gerlach separated site-resolved band mapping images (\fig[a]{fig2}). 
The amplitude~$A$ was found by fitting 
\begin{align}
\mathcal{I}_{\text{fit}}\left(t^\prime\right) = &A\,e^{-t^\prime/\tau_{\text{ex}}}\sin\left(\omega_{\text{ex}} t^\prime + \theta\right) + \mathcal{I}_{\text{s}} + \mathcal{I}_\text{d}\,e^{-t^\prime/\tau_\text{d}}
\label{eq:fitfunction}
\end{align}
to the oscillation data, where $t^\prime=t-t_\text{s}$ and $\theta\approx 0$ a phase shift induced by a finite freezing ramp speed.
Compared to the ideal evolution, two additional effects are taken into account.
First, an exponential decay of the amplitude with a time constant~$\tau_\text{ex}$ accounts for dephasing between individual double wells.
Both, $\tau_\text{ex}$ and the oscillation frequency~$\omega_\text{ex}\simeq (E_2-E_1)/\hbar$ were determined for each~$\Delta_\text{s}$ with an independent superexchange oscillation measurement.
Second, an additional decay of the imbalance offset~$\mathcal{I}_\text{d}$ is caused by an exponential relaxation of a small residual magnetic field gradient after the abrupt stop of $B^\prime=-24.3(6)\,\text{Hz}/d_s$, with a decay constant $\tau_\text{d}=1.05(5)\,\text{ms}$.
The resulting oscillation amplitudes during the pump cycle for two different~$J_\text{ex}$ as well as exemplary spin imbalance traces are summarized in \fig[b]{fig2}.
The spin current peaks around $\Delta=0$, where the ground state is delocalized and spins move.
For large gradients the eigenstates are independent of $\Delta$ and the spin current vanishes.
Note that the residual gradient $\Delta_d$ slightly shifts the peak towards higher $\Delta$.
Furthermore, the spin current strongly depends on the exchange coupling~$J_\text{ex}$.
With increasing $J_\text{ex}$, the wavefunction delocalizes and depends less on~$\Delta$, so the peak width increases while the maximum amplitude decreases.
However, unlike the instantaneous current, the transported spin during one pump cycle, is independent of the pump parameters.

\begin{figure}%
\includegraphics[width=\columnwidth]{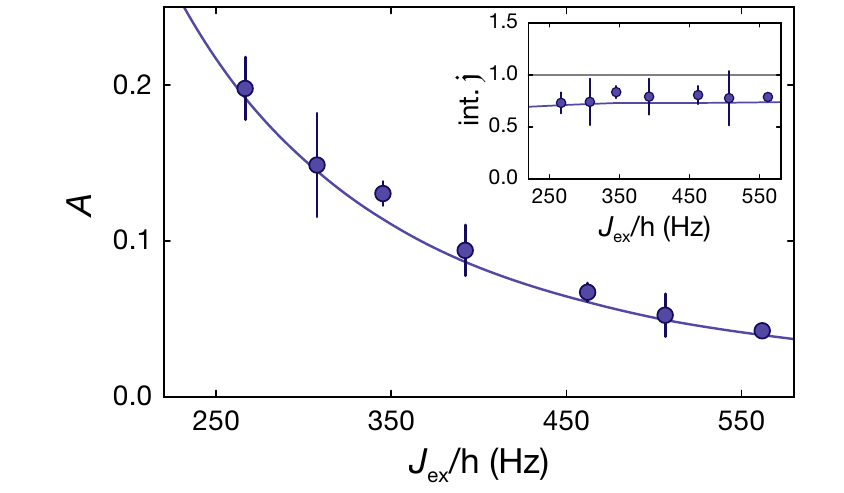}%
\caption{%
Imbalance oscillation amplitude~$A$ at $\Delta_\text{s}=0$ as a function of~$J_\text{ex}$.
Each point is an average of the fitted amplitudes of three time traces; the error bars are the standard deviations.
The blue lines are a numerical calculation taking into account the reduced detection efficiency as well as the measured initial state occupation and finite pump efficiency.
Assuming a constant scaling factor for each~$J_\text{ex}$ as indicated by the measurement in \fig{fig2}, the integrated current per cycle can be estimated and is shown in the inset.
In the ideal case, the integrated current is equal to one (gray line).
}%
\label{fig3}%
\end{figure}
\begin{figure}%
\includegraphics[width=\columnwidth]{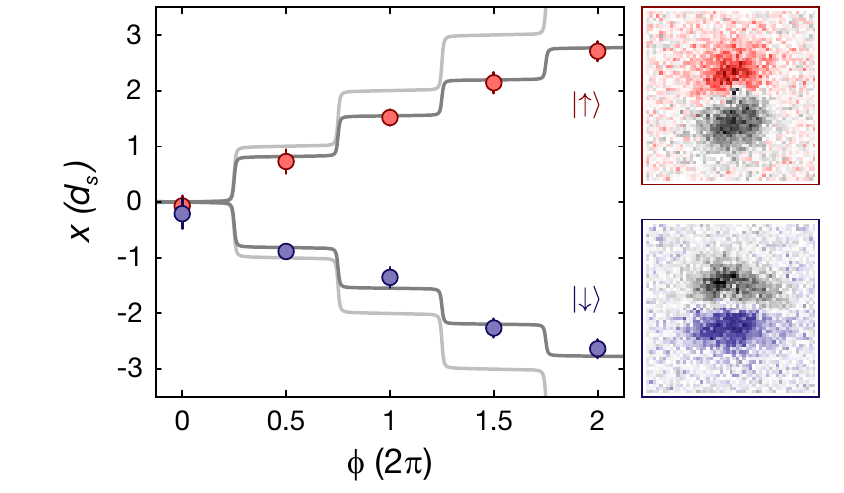}%
\caption{%
Center-of-mass position of up (red) and down (blue) spins as a function of the pump parameter~$\phi$.
The points show the center-of-mass position averaged over ten data sets of a spin-selective imaged atom cloud; the error bars show the error of the mean.
Each data set consists of an average of ten pairs, which contain an image obtained by a sequence with pumping and one using a reference sequence with the same length but constant pump parameter~$\phi=0$.
Difference images of both sequences for up and down spin are shown on the right side.
The solid lines depict the calculated motion of a localized spin for the ideal case (light gray) and for a reduced ground state occupation and a pump efficiency per half pump cycle that was determined independently through a band mapping sequence (gray). 
}%
\label{fig4}%
\end{figure}

\invisiblesection{Numerical model}
To compare the data with theoretical expectations, we performed a numerical calculation including a reduced detection efficiency caused by the residual gradient decay.
Imbalance time traces for $\psi_{t_s}$  were evaluated using a two spin, two-site extended Bose-Hubbard Hamiltonian $\hat{H}_\text{DW}(t^\prime)$ with $\Delta(t^\prime)=\Delta_\text{d}\,e^{-t^\prime/\tau_\text{d}}+\Delta_\text{s}$~\cite{Scarola:2005,Supplementary}.
The calculated time traces were also fitted with \eq{eq:fitfunction}; the resulting oscillation amplitude describes ideal transport of ground state spins (light lines in \fig{fig2}). 
This curve can be fitted to the data by rescaling the amplitude with a factor of $0.84(6)$, which directly determines the reduction of the integrated measured current compared to the ideal one and thereby the transported spin polarization.
The deviation from ideal transport can be attributed to an imperfect initial state preparation with $97(1)\,\%$ ground state occupation and a pump efficiency per half a pump cycle of $89(1)\,\%$, which describes the fraction of double wells that remain in the ground state after half a cycle.
Considering this additional occupation of the first excited state, which creates an opposite current, the total spin current can be calculated (dark lines in \fig[b]{fig2}).
Fitting these expected oscillation amplitudes to the measured data by rescaling with a global factor results in a fit value of $1.05(8)$ for $J_\text{ex}/h=342(2)\,\text{Hz}$ and $1.06(8)$ for $J_\text{ex}/h=467(3)\,\text{Hz}$.
This shows that even though the shape and amplitude of the curve changes, the integrated current is only defined by the pumps' topology not by the specific tunneling parameters.
Furthermore, we note that the deviation to the theoretically expected integrated current can not be attributed to edge effects as they are negligibly small for the present trapping potential.

To study the dependence of the maximum current on the exchange coupling, the oscillation amplitude was measured at $\Delta_\text{s}=0$ for various~$J_\text{ex}$ (\fig{fig3}). 
The maximum amplitude decreases with rising~$J_\text{ex}$, as the spins' wavefunction is more delocalized for the same~$\Delta$ and therefore the current flow is spread over a larger sector in the pump cycle.
The observed peak amplitude agrees with the numerical model including the initial ground state occupation and pump efficiency.
As suggested by the measurements in \fig{fig2}, the integrated spin current can be extracted by rescaling the ideal amplitude with a global factor and is found to be constant for all exchange couplings (inset \fig{fig3}).

\invisiblesection{In-situ spin transport}
Independent evidence for the spin separation and a quantitative comparison with the total spin current can be obtained by measuring the center-of-mass position of the two spin components from in-situ absorption images after removing one of the spin components.
When varying~$\phi$, the up and down spins clearly separate (\fig{fig4}).
This independently verifies the spin transport and shows quantitative agreement with the results of the spin current measurement~\cite{Supplementary}.

\invisiblesection{Conclusions}
In conclusion, we have demonstrated the implementation of a spin pump and introduced a new method for directly measuring instantaneous spin currents.
Comparing the measured spin imbalance oscillation amplitudes with the adiabatic theory shows that the integrated current is independent of the specific pump parameters and gives evidence for the utility of the developed current measurement method.
The method can also be extended to more general systems by performing an instantaneous projection onto double wells.
Investigating such spin pumps on a single site level would allow for local observation of spin currents and the direct observation of edge excitations in finite systems~\cite{Hatsugai:2016}.
A system described by the non-trivial $Z_2$-invariant can be realized with time-reversal invariant spin orbit interaction~\cite{Shindou:2005,Zhou:2014}.
When breaking time-reversal symmetry, the topological properties of the QSH system remain but spin-Chern numbers are required for the description~\cite{Zhou:2014}.
Furthermore, a topological, interaction-driven quantum motor~\cite{Zhou:2004, Bustos-Marun:2013} can be accomplished by only pumping one of the components while the other is coupled by interaction.
For spin pumps with highly degenerate many-body ground states, fractional transport is predicted~\cite{Meidan:2011}.

\invisiblesection{Acknowledgments}
We acknowledge insightful discussions with M.~Aidelsburger. 
This work was supported by NIM, the EU (UQUAM, SIQS), and the DFG (DIP \& FOR2414). M.\,L. was additionally supported by ExQM and R.\,C. by FIRB-2012-HybridNanoDev (Grant No. RBFR1236VV).

\bibliographystyle{bosons}

\renewcommand{\thefigure}{S\the\numexpr\arabic{figure}-10\relax}
 \setcounter{figure}{10}
\renewcommand{\theequation}{S.\the\numexpr\arabic{equation}-10\relax}
 \setcounter{equation}{10}
 \renewcommand{\thesection}{S.\Roman{section}}
\setcounter{section}{10}
\renewcommand{\bibnumfmt}[1]{[S#1]}
\renewcommand{\citenumfont}[1]{S#1}

\onecolumngrid
\clearpage
\invisiblesection{Supplemental Material}
\begin{center}
\noindent\textbf{Supplemental Material for:}
\\\bigskip
\noindent\textbf{\large{Spin pumping and measurement of spin currents in optical superlattices}}
\\\bigskip
C.~Schweizer$^{1,2}$, M.~Lohse$^{1,2}$, R.~Citro$^{3,4}$, I. Bloch$^{1,2}$
\\\vspace{0.1cm}
\small{$^1$\,\emph{Fakult\"at f\"ur Physik, Ludwig-Maximilians-Universit\"at, Schellingstrasse 4, D-80799 M\"unchen, Germany}}\\
\small{$^2$\,\emph{Max-Planck-Institut f\"ur Quantenoptik, Hans-Kopfermann-Strasse 1, D-85748 Garching, Germany}}\\
\small{$^{3}$\,\emph{Dipartimento di Fisica ``E.~R.~Caianiello'', Università degli Studi di Salerno,\\
Via Giovanni Paolo II 132, I-84084 Fisciano (Salerno), Italy}}\\
\small{$^{4}$\,\emph{SPIN-CNR Salerno, Via Giovanni Paolo II 132, I-84084 Fisciano (Salerno), Italy}}
\end{center}
\bigskip
\bigskip
\twocolumngrid

\newcommand{\newparagraph}{~\\\vspace{-2em}\paragraph}

\section{Spin current measurement}
\subsection{Two-site extended Bose-Hubbard model}
In the tight binding limit, two spins on an isolated double well ($\delta J = J$) can be described by a two-site Bose-Hubbard model, where we denote the spins $\sigma=\left\{{\uparrow}, {\downarrow}\right\}$ and the left (right) site with L (R).
The Hamiltonian is given by
\begin{equation}
\begin{aligned}
\hat{H}\sub{BH}=&-J\sum_{\sigma=\left\{\uparrow,\downarrow\right\}}
\left(\hat{a}^{\dagger}_{\text{L},\sigma}\hat{a}^{\phantom\dagger}_{\text{R},\sigma}
+\text{h.c.}\right)\\
&-\frac{\Delta}{2}\left(\hat{n}_{\text{L},\downarrow}-\hat{n}_{\text{L},\uparrow}-\hat{n}_{\text{R},\downarrow}+\hat{n}_{\text{R},\uparrow}\right)\\
&+U\left(\hat{n}_{\text{L},\uparrow}\hat{n}_{\text{L},\downarrow}+\hat{n}_{\text{R},\uparrow}\hat{n}_{\text{R},\downarrow}\right)
\label{eq:suppl:BoseHubbard}
\end{aligned}
\end{equation}
with~$\hat{a}^{\dagger}_{\text{R/L},\sigma}$ ($\hat{a}^{\phantom\dagger}_{\text{R/L},\sigma}$) the creation (annihilation) operator of spin~$\sigma$ on the left or right site, $\hat{n}_{\text{R/L},\sigma}$ the number operator counting the spins, $J$~the tunneling rate, $\Delta$ the spin-dependent energy offset between left and right site, and $U$ the on-site interaction energy.
The accuracy of the exchange coupling, especially at large $J/U$, can be increased by including corrections from density-dependent hopping $J_\text{ddh}=-g\int w^3\sub{L}(r) w\sub{R}(r)\,\text{d}^3r$ and nearest neighbor interaction $U_\text{LR}=g\int w^2\sub{L}(r) w^2\sub{R}(r)\,\text{d}^3r$~\cite{Scarola:2005:Suppl, Trotzky:2007:Suppl}.
Here, $w_\text{L/R}(r)$ denotes the Wannier function on the left/right site.
The Hamiltonian of this extended Bose-Hubbard model~$\hat{H}_\text{DW}$ is in the basis $\ket{\uparrow\downarrow,0}$, $\ket{\uparrow,\downarrow}$, $\ket{\downarrow,\uparrow}$, $\ket{0,\uparrow\downarrow}$ with $J^\prime=J+J_\text{ddh}$:
\begin{equation}
\hat{H}_\text{DW}=\begin{pmatrix}
	U						&-J^\prime						&-J^\prime						&U_\text{LR}\\
	-J^\prime		&U_\text{LR}+\Delta	&U_\text{LR}					&-J^\prime	\\
	-J^\prime		&U_\text{LR}					&U_\text{LR}-\Delta	&-J^\prime	\\
	U_\text{LR}	&-J^\prime						&-J^\prime						&U
\end{pmatrix}.
\label{eq:suppl:HMatrix}
\end{equation}
The dependence of the energy spectrum~$E_n$ of the extended Bose-Hubbard model on~$\Delta$ is shown in \fig[a]{fig:suppl:bandstructureandamcoeffs}.
We define the exchange coupling~$J_\text{ex}$ as the gap between the first and second eigenenergy at $\Delta=0$.
The ground state in the limit~$\left|\Delta\right|\gg J_\text{ex}$ is $\left|1\right>\approx\left|\downarrow,\uparrow\right>$ for positive and $\left|\uparrow,\downarrow\right>$ for negative $\Delta$.
At $\Delta=0$ and $U\gg J$ the ground state is approximately a triplet state $\frac{1}{\sqrt{2}}\bigl(\left|\uparrow,\downarrow\right>+\left|\downarrow,\uparrow\right>\bigr)$.
Note that the third eigenstate is independent of~$\Delta$.

\begin{figure}%
\includegraphics[width=\columnwidth]{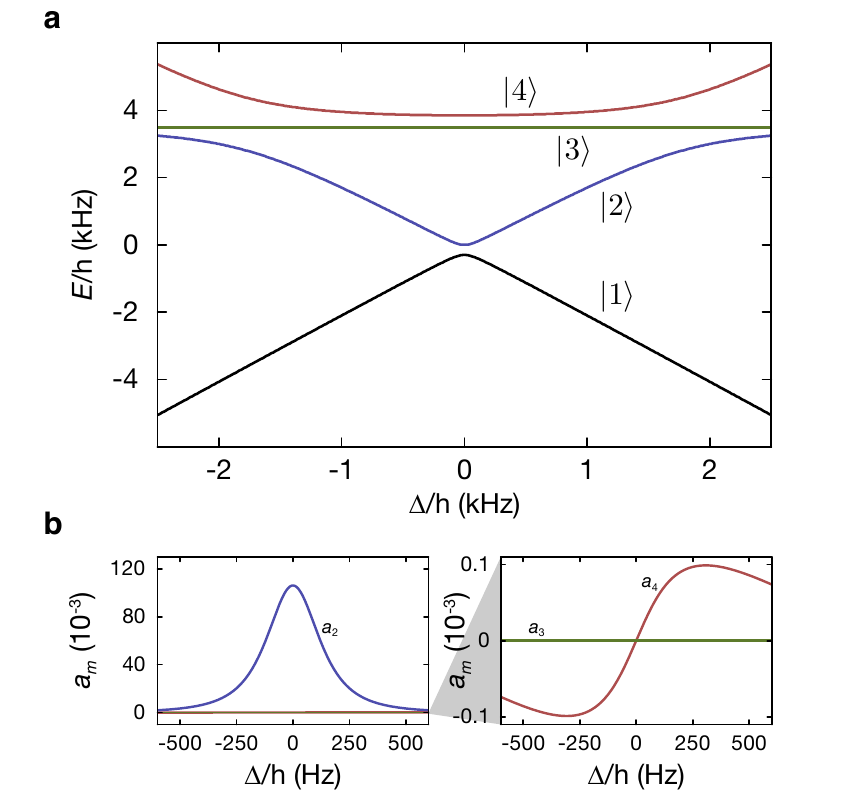}%
\caption{Energy spectrum and $a_m$-coefficients. (a)~Energy spectrum of the extended Bose-Hubbard model for two spins on a double well. (b)~Magnitude of the $a_m$-coefficients at different spin-dependent tilt values~$\Delta$ during the magnetic field ramp with a speed of $\dot{\Delta}=82(2)\,\text{kHz/s}$ and exchange coupling of $J_\text{ex}/h=342(2)\,\text{Hz}$.}%
\label{fig:suppl:bandstructureandamcoeffs}%
\end{figure}

An adiabatic change of the tilt~${\Delta}$ with rate~$\dot{\Delta}$ leads in first-order approximation to the temporary wavefunction 
\begin{equation}
\ket{\psi_t}=\ket{n_t}+\ii\sum_{m\neq n}a_{m}(t)\ket{m_t}
\label{eq:suppl:generaltemporarywavefunction}
\end{equation}
using parallel transport conditions \cite{Xiao:2010:Suppl}. 
Here, $\left|l_t\right>$ is an instantaneous eigenstate of $\hat{H}_\text{DW}$ and the  admixture coefficients are
\begin{equation}
a_m(t)=-\frac{\bra{m_t}\hbar\partial_{t}\ket{n_t}}{E_n(t)-E_m(t)}=-\dot{\Delta}\frac{\bra{m_t}\hbar\partial_{\Delta}\ket{n_t}}{E_n(t)-E_m(t)}.
\label{eq:suppl:amcoeffs}
\end{equation}
A detailed derivation can be e.g.\ found in Ref.\,\cite{Xiao:2010:Suppl}.
The $a_m$-coefficients for $\left|n_t\right>=\left|1_t\right>$, i.e.\ starting from the ground state, are depicted in \fig[b]{fig:suppl:bandstructureandamcoeffs}; the contribution~$a_2$ clearly dominates and other coefficients can be neglected because the dependence of the eigenstates $\left|4\right>$ and $\left|3\right>$ on $\Delta$ is weak or even vanishing because of the large gap $|E_m-E_1| \sim U\gg J_\text{ex}$, for $m>2$.
Then the temporary wavefunction~\eq{eq:suppl:generaltemporarywavefunction} reduces to
\begin{equation}
\ket{\psi_t}\approx\ket{1_t}+\ii\,a_{2}(t)\ket{2_t}.
\label{eq:suppl:temporarywavefunction}
\end{equation}

\subsection{Connection Between Amplitude and Current}
For the current measurement, the pump cycle is abruptly stopped at~$\Delta_\text{s}$.
In the ideal case this happens instantaneously, such that the wavefunction~$\left|\psi(\Delta_\text{s})\right>$ subsequently evolves according to the static Hamiltonian~$\hat{H}_\text{DW}(\Delta_\text{s})$.
Since $\left|\psi(\Delta_\text{s})\right>$ is in general not an eigenstate of $\hat{H}_\text{DW}(\Delta_\text{s})$, the expectation value of the imbalance operator~$\mathcal{I}$ oscillates in time
\begin{equation}
\mathcal{I}(t^\prime)=A\,\sin\left(\frac{E_2(t_\text{s})-E_1(t_\text{s})}{\hbar}~t^\prime\right)+\mathcal{I}_\text{s}
\label{eq:suppl:imboscillation}
\end{equation}
with $t^\prime=t-t_\text{s}$. The oscillation amplitude is directly proportional to the $a_2$-coefficient
\begin{equation}
A=-2\,a_2(t_\text{s})\left<1_{t_\text{s}}\right|\hat{\mathcal{I}}\left|2_{t_\text{s}}\right>.
\label{eq:suppl:amplitude}
\end{equation}
When comparing this result to the the integral form of the continuity equation, the spin current at~$t_\text{s}$ is directly connected to the oscillation amplitude~$A$ via
\begin{equation}
\begin{aligned}
\mathfrak{j}(t_\text{s})&=\left.\frac{1}{2}\partial_{t}\mathcal{I}(t^\prime)\right|_{t=t_\text{s}}\\
&=A\,\frac{E_2(t_\text{s})-E_1(t_\text{s})}{2\hbar}\,\left.\cos\left(\frac{E_2-E_1}{\hbar}~t^\prime\right)\right|_{t=t_\text{s}}\\
&=A\,\frac{E_2(t_\text{s})-E_1(t_\text{s})}{2\hbar}.
\label{eq:suppl:currentcontinuityequation}
\end{aligned}
\end{equation}

\subsection{Oscillation Amplitude -- Model and Corrections}
\begin{figure}%
\includegraphics[width=\columnwidth]{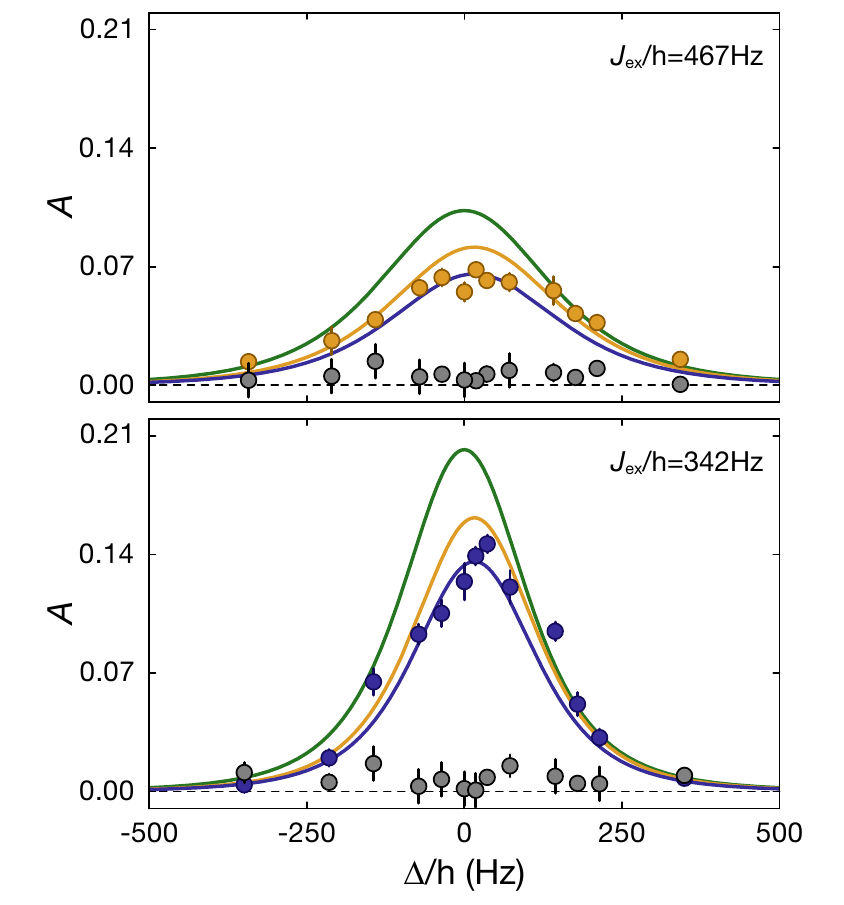}
\caption{Current measurement. The figure summarizes different theory curves for both data sets at $J_\text{ex}/h=467(3)\,\text{Hz}$ (orange) and $J_\text{ex}/h=342(2)\,\text{Hz}$ (blue) from \ref{fig2} in the main text. %
The green solid line shows the ideally expected oscillation amplitude.
A corrected description taking into account the reduced detection efficiency due to a residual exponential decay of the magnetic field gradient (orange solid line), and a description including additionally a finite ground state occupation of $97(1)\,\%$ and a pump efficiency of $89(1)\,\%$ (blue solid line) are shown. Furthermore, a measurement of the particle current is shown (gray), obtained by analyzing the particle imbalance oscillations between the left and right side of the double well based on the same data sets as for the spin imbalance oscillations.}%
\label{fig:suppl:currentdetails}%
\end{figure}
For two spins occupying the ground state of an isolated double well, the ideal oscillation amplitude for a perfectly abrupt stop of the ramp can be evaluated from \eq{eq:suppl:amplitude}.
This ideal amplitude, within~$\hat{H}_\text{DW}$, is shown as a function of $\Delta_\text{s}$ in \fig{fig:suppl:currentdetails} (green solid lines).
As the abrupt stop is smoothed by a decaying residual magnetic gradient, the measured amplitude for a given spin current is slightly reduced.
This amplitude can be calculated numerically by solving the time-dependent Schrödinger equation including the residual magnetic gradient decay.
The model assumes an initial state after the adiabatic evolution (\eq{eq:suppl:temporarywavefunction}) stopped at $\Delta_\text{s}+\Delta_\text{d}$  and a time-dependent Hamiltonian $\hat{H}_\text{DW}(\Delta_\text{d} e^{-t^\prime/\tau_\text{d}}+\Delta_\text{s})$.
For long times $t^\prime\gg\tau_\text{d}$, the Hamiltonian approaches the ideal case.
With a numerical time propagation, the evolution of the wavefunction and thereby $\mathcal{I}(t^\prime)$ can be calculated.
From this time trace the amplitude is extracted with a fit of \ref{eq:fitfunction} from the main text as for the experimental data.
The effect of this detection correction is visualized in \fig{fig:suppl:currentdetails} as orange solid lines.

Preparing only ground state double wells is extremely challenging and therefore a small residual amount of excited state double wells is expected. 
When considering only ground and first excited state, the spin imbalance is a good measure for the state occupation at $\left|\Delta\right|\gg J_\text{ex}$, where the spins are perfectly localized.
By measuring the spin imbalance before~($\mathcal{I}_\text{i}$) and after half a pump cycle~($\mathcal{I}_\text{f}$), the initial state preparation and the pump efficiency can be characterized.
The pump efficiency~$\beta_1$ is a measure of the fraction of double wells that remain in the ground state during half a pump cycle and can be deduced from the spin imbalance: $\beta_1=\frac{\mathcal{I}_\text{i}-\mathcal{I}_\text{f}}{2\mathcal{I}_\text{i}}$.
The initial fraction of ground and excited state double wells is $n_1^{\text{(i)}}=\frac{\mathcal{I}_1+\mathcal{I}_\text{i}}{2\mathcal{I}_1}$ and $n_2^{\text{(i)}}=\frac{\mathcal{I}_2+\mathcal{I}_\text{i}}{2\mathcal{I}_2}=\frac{\mathcal{I}_1-\mathcal{I}_\text{i}}{2\mathcal{I}_1}$, assuming $\mathcal{I}_1=-\mathcal{I}_2$ being the ideal spin imbalance at the initial conditions.

The integrated spin current~$\overline{\mathfrak{j}}=\int_{t_\text{i}}^{t_\text{f}}\mathfrak{j}(t)\,\text{d}t$ is the sum of the ideal integrated spin currents $\overline{\mathfrak{j}}_{1/2\text{,ideal}}$ of both states weighted by the initial band occupation and the pump efficiency.
The ideal spin currents are oppositely directed $\overline{\mathfrak{j}}_{1\text{,ideal}}=-\overline{\mathfrak{j}}_{2\text{,ideal}}=\mathcal{I}_1$ and given by the ideal initial spin imbalance.
Furthermore, equal pump efficiency $\beta_2=\beta_1$ for ground and first excited state are assumed.
\begin{equation}
\begin{aligned}
\overline{\mathfrak{j}}&=n_1^{\text{(i)}}\beta_1\,\overline{\mathfrak{j}}_{1\text{,ideal}}+n_2^{\text{(i)}}\beta_2\,\overline{\mathfrak{j}}_{2\text{,ideal}}\\
 &=\left(n_1^{\text{(i)}}-n_2^{\text{(i)}}\right)\beta_1\,\overline{\mathfrak{j}}_{1\text{,ideal}}\\
 &=\frac{\mathcal{I}_\text{i}-\mathcal{I}_\text{f}}{2}
\end{aligned}
\label{eq:suppl:idealtotalspincurrent}
\end{equation}
The result can be verified by a comparison with the integral form of the continuity equation $2\mathfrak{j}=\partial_t\mathcal{I}$.

The reduction of the integrated current as a result of a finite excited state occupation and a reduced pump efficiency can be approximately captured in the data analysis by rescaling of the current~$\mathfrak{j}(t)$ with a global factor.
Such a rescaling corresponds to a description by an average state occupation with perfect pump efficiency.
The blue line in \fig{fig:suppl:currentdetails} shows the spin current taking into account the detection efficiency, the initial ground state occupation as well as the pump efficiency.

\section{Energy spectrum}
\begin{figure}%
\includegraphics[width=\columnwidth]{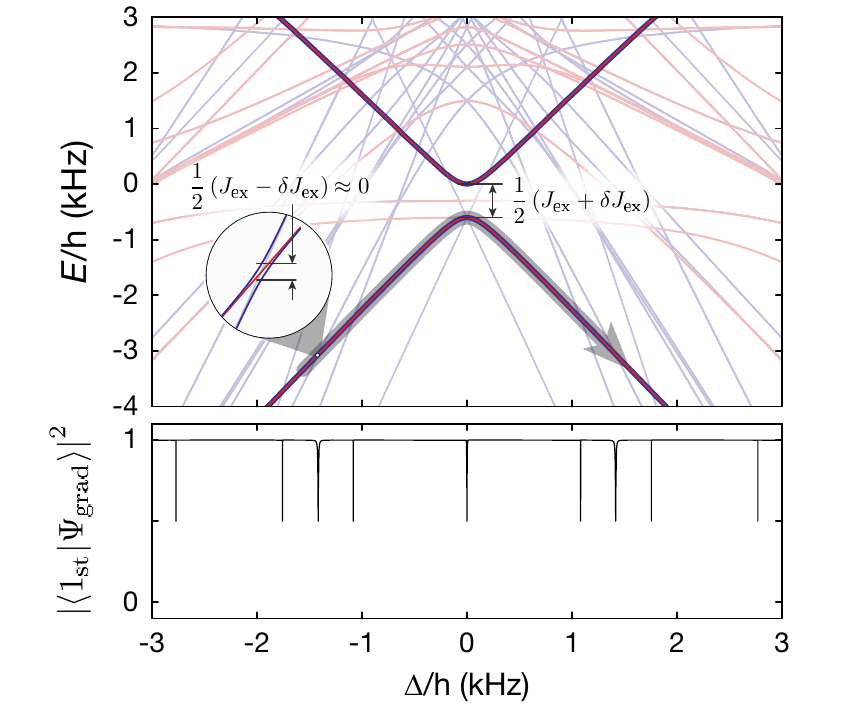}%
\caption{Energy spectrum and state overlap for a superlattice with staggered tilt and a global gradient, respectively. In the upper panel the energy spectrum of two up and two down spins on two double wells with local, staggered tilt (red) and global gradient (blue) is shown. The thicker darker lines represent the ground state of the staggered superlattice $\left|1_\text{st}\right>$ and the corresponding state in the globally tilted lattice  $\left|\phi_\text{grad}\right>$, which is used for pumping in the experiment.  In the lower panel the state overlap between these two states is depicted.}%
\label{fig:suppl:spectrum}%
\end{figure}

In the experiment, the spin chain with local tilts is realized in the limit of isolated double wells with a global magnetic gradient. 
In this system unlike for $\hat{H}$ in \ref{eq:spinmodel} of the main text, the prepared initial state is not the ground state but an excited state.
During the pump cycle, however, it evolves in the same way as the ground state of a spin-dependent superlattice with the exception of a number of additional crossings that occur in the energy spectrum (see \fig{fig:suppl:spectrum}).
However, the gaps are very small -- on the order of the inter double well exchange coupling $\frac{1}{2}(J_\text{ex}-\delta J_\text{ex})$ -- and can be crossed non-adiabatically.
In conclusion the gradient model requires not only adiabaticity with respect to the intra double well exchange tunneling $\frac{1}{2}(J_\text{ex}+\delta J_\text{ex})$ but also pure non-adiabatic transfer with respect to gaps on the order of the inter double well exchange coupling $\frac{1}{2}(J_\text{ex}-\delta J_\text{ex})$.
As the latter can be suppressed exponentially with the long lattice depth, and adiabaticity on the double well scale can be reached by slower ramp speed, the transport in these models can be described by the same topological invariant.
An energy spectrum for two up and two down spins on two double wells is shown in \fig{fig:suppl:spectrum} for the experimental parameter set for $J_\text{ex}/h=342(2)\,\text{Hz}$.
The pump cycle follows the thicker darker depicted state, which is either strongly gapped around $\Delta=0$ or crosses states with negligibly small gaps. 
Additionally, the state overlap between the groundstate in the staggered model and the pumped state for the model with a global gradient is calculated and found to be one, apart from the vicinity of the tiny gaps.
This shows that the pumped state in the gradient model is very similar to the one in the model with staggered tilts.

\section{Experimental sequence}
\begin{figure}%
\includegraphics[width=\columnwidth]{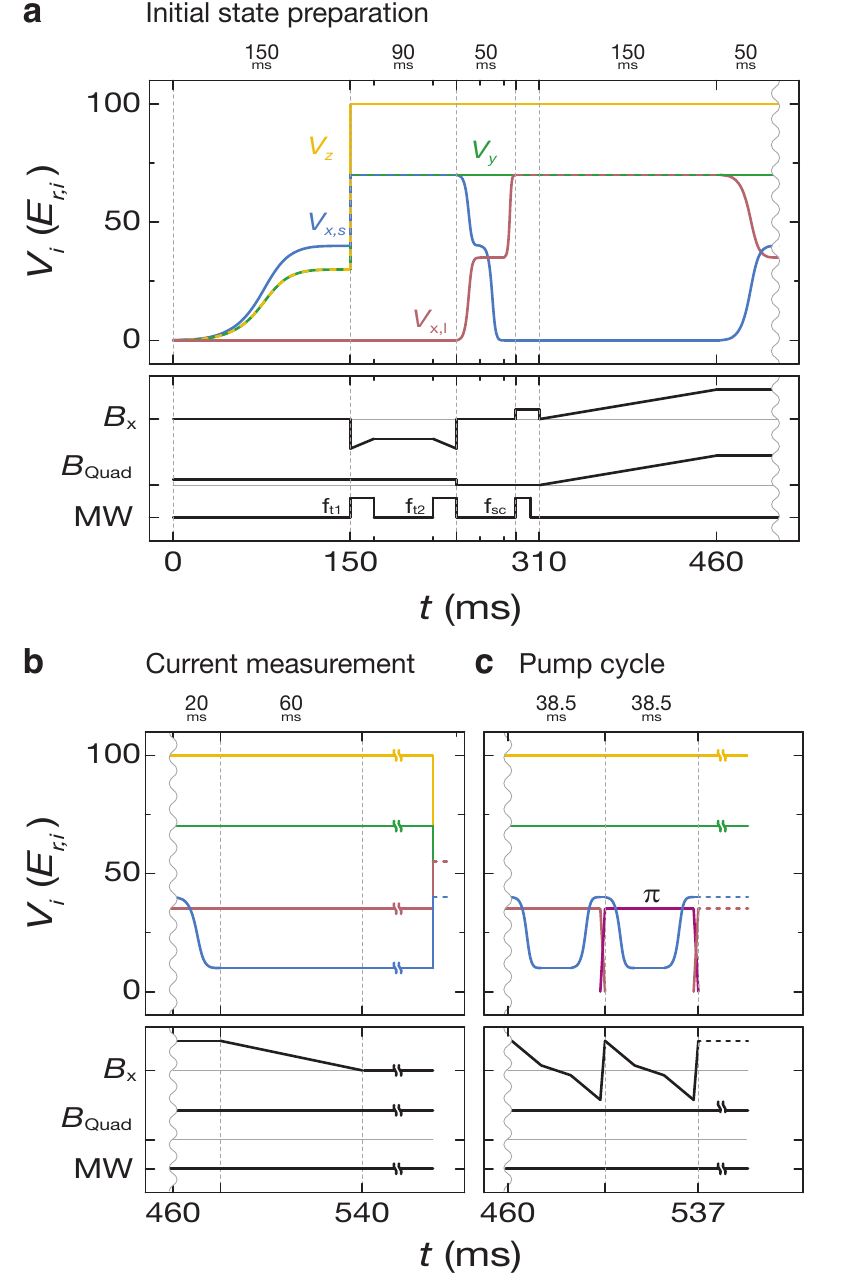}
\caption{Summary of the experimental sequences. The lattice depths are shown in each upper panel in their respective recoil energies: along the $x$-direction with period~$d_s$ (blue) and~$d_l$ (red), along the $y$-direction with period~$d_s$ (green) and along the $z$-direction with period~$d_z$ (yellow). In the lower panel the two contributions to the magnetic field gradient as well as the mircowave pulses are sketched. 
(a)~A time sequence for the initial state preparation is depicted. It ends with decoupled sites occupied by antiferromagnetically ordered $\ket{\uparrow}=\ket{F=1,m_{F}=-1}$ and $\ket{\downarrow}=\ket{F=1,m_{F}=+1}$ spins. 
(b)~Example sequence for the current measurement at $\Delta_\text{s}=0$ which directly follows the initial state preparation. 
(c)~The spin pump sequences for the in-situ and band mapping measurement starting directly after the initial state preparation. Note after half a cycle the dimerization is flipped by swapping to a second laser creating $\pi$ shifted double wells, and the magnetic field gradient is reversed in $2\,\text{ms}$.}%
\label{fig:suppl:sequence}%
\end{figure}

The experimental sequence starts with a Mott insulator of $(F=1,m_{F}=-1)$ $^{87}$Rb atoms in a 3D optical lattice of three mutually orthogonal standing waves with wavelengths $\lambda_{x}=\lambda_{y}=767$\,nm and $\lambda_{z}=844$\,nm. 
With a sequence of microwave driven adiabatic passages, the atoms are transferred to the $(F=1,m_{F}=0)$ via the $(F=2,m_{F}=+1)$ state (see $f_\text{t1}$, $f_\text{t2}$ in \fig{fig:suppl:sequence}). 
Then two neighboring lattice sites are merged along the $x$-direction into decoupled sites with twice the period by ramping up a long lattice with period $d_\text{l}=2d_\text{s}$, where $d_\text{s}=\lambda_{x}/2$, and simultaneously turning the short period lattice off.
With coherent microwave-mediated spin changing collisions~\cite{Widera:2005:Suppl} each atom pair is transferred to a pair of $\ket{\uparrow}=\ket{F=1,m_{F}=-1}$ and $\ket{\downarrow}=\ket{F=1,m_{F}=+1}$ atoms. 
Subsequently, a magnetic field gradient is turned on and the long lattice sites are split adiabatically into two decoupled sites by turning on the short period lattice to a depth of $V_\text{s}=40\,{E}_\text{r,s}$ in $50$\,ms with ${E}_\text{r,i}=\frac{h^2}{2m_\text{Rb}\lambda_\text{i}^2}$, where $m_\text{Rb}$ is the mass of a Rubidium atom.
Due to the present magnetic field gradient of $\Delta/h=2.7(2)\,$kHz during the splitting, the up and down spins order antiferromagnetically in individual, decoupled sites with $J_\text{ex}\approx 0$.
A detailed time sequence of the individual experimental parameters is shown in \fig[a]{fig:suppl:sequence}.
This initial state preparation is then followed by the pumping sequence.

The spin pump sequence for multiple cycles, which was used for the in-situ measurements, starts by coupling neighboring lattice sites of each double well by decreasing the short period lattice to the final value for the corresponding superexchange coupling.
Then, the spin-dependent gradient is inverted adiabatically by changing the bias field along the $x$-direction (see gradient calibration), and at the end of the first half pump cycle the distribution is frozen again by increasing the short lattice depth such that $J_\text{ex}\approx 0$.
To continue the cycle, the dimerization is changed by swapping to a second laser which creates a $\pi$-shifted long lattice, and simultaneously setting the magnetic bias field back to its initial value in $2\,\text{ms}$.
Then the previously described cycle is repeated multiple times.
Before in-situ imaging, residual atoms in the ($F=1$, $m_F=0$) state are removed by applying simultaneously a microwave field on the  ($F=1$, $m_F=0$)$\rightarrow$($F=2$, $m_F=0$) transition and a resonant imaging light pulse to remove the atoms in the $F=2$ manifold.
Subsequently, one of the spin states is selected and transferred to the ($F=2$, $m_F=0$) state by a microwave driven adiabatic passage and imaged by absorption imaging.
 
The current measurement sequence is limited to the first half pump cycle.
Neighboring sites are coupled by decreasing the short lattice depth $V_\text{s}$ and the bias field~$B_x$ is changed with a constant rate to different final values of the magnetic tilt~$\Delta_\text{s}$, where it is abruptly stopped. 
After a variable holdtime~$t$, the spin distribution is frozen by ramping up the short period lattice and a site-resolved band mapping was performed.
During time-of-flight, the spin-components are separated spatially by a Stern-Gerlach field and the four site occupations $n_{\text{L}\downarrow}$, $n_{\text{L}\uparrow}$, $n_{\text{R}\downarrow}$, and $n_{\text{R}\uparrow}$ can be simultaneously extracted from each absorption image.

\section{Calibration of lattice, gradient field and Hubbard parameters}
\newparagraph{Calibration of bare tunneling rate:}
The bare tunneling matrix element~$J$ couples two neighboring sites and can be calibrated from left-right oscillations of a single atom on a double well with degenerate on-site energies ($\Delta = 0$).
At the beginning, a single atom is prepared on the left site of each double well.
To this end, an $n=1$~Mott insulator in the long lattice is prepared and adiabatically split with the short lattice in the presence of a large potential tilt.
The short lattice depth is increased such that the atoms localize on the lower-lying left site ($J\approx 0$).
Then, the tilt is removed and subsequently the short period lattice is ramped down in $200\,\upmu\text{s}$ to the lattice depth at which~$J$ needs to be calibrated.
The localized wavefunction is not an eigenstate of the symmetric double well, but an equal superposition of the ground~$\left|1\right>$ and first excited state~$\left|2\right>$ and thus the left-right occupation oscillates in time.
The left-right fraction can be measured with site-resolved band mapping and its oscillation frequency~$f_\text{bare}$ is equal to the difference of the single-particle eigenenergies $hf_\text{bare}=\epsilon_{2}-\epsilon_{1}=2J$.

\newparagraph{Calibration of on-site interaction energy:}
The on-site interaction energy~$U$ is the extra energy for placing a second particle on the same site and can be calibrated via the superexchange oscillation frequency. 
An up and a down spin are localized on the left and right site of a double well prepared in the presence of a large spin-dependent tilt $\Delta$ as for the current measurement.
Then, the magnetic fields, except for a small bias field to maintain the quantization axis, are switched off to non-adiabatically remove the tilt while $J_\text{ex}\approx 0$ and afterwards the short period lattice is decreased within $200\,\upmu\text{s}$ to the final value, at which $U$ is calibrated. 
The localized state is not an eigenstate anymore and evolves in time.
The time evolution leads to superexchange oscillations \cite{Trotzky:2007:Suppl}, which for the experimentally used parameters are dominated by the contributions from the ground and first excited state $\hbar \omega_\text{ex}\simeq E_2-E_1=J\sub{ex}$.
The exchange coupling~$J_\text{ex}$ can be calculated from an extended Bose-Hubbard model and depends strongly on~$U$.
Thus, $U$ can be inferred from~$\omega_\text{ex}$.

\newparagraph{Gradient Calibration:}
\begin{figure}%
\includegraphics[width=\columnwidth]{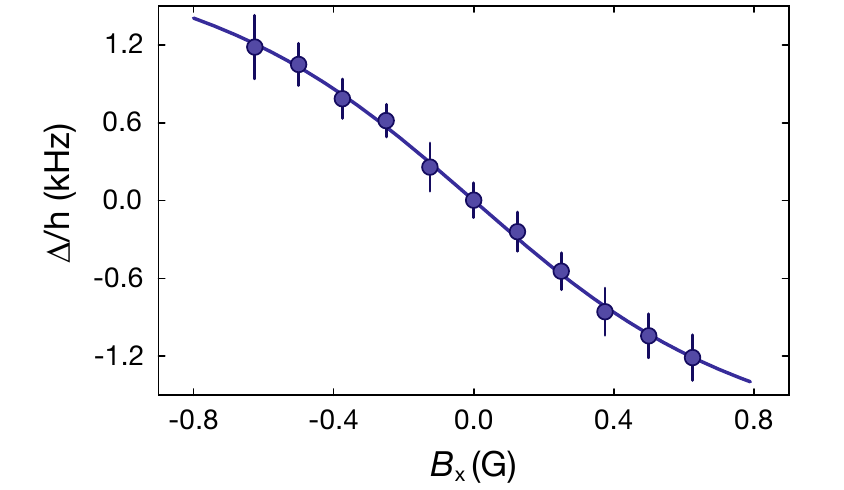}%
\caption{Calibration of the magnetic gradient. The data points show a measurement of the magnetic tilt~$\Delta$ by laser-assisted tunneling spectroscopy in a tilted double well as a function of the magnetic bias field~$B_x$. From the measured frequency, the tilt can be directly inferred by $2\Delta=h\,f_\text{r}-\delta_0$. The error bars show the standard deviation of an average of eight points, four for ($F=2$, $m_F=\pm2$) each.}%
\label{fig:suppl:bfieldcalib}%
\end{figure}
The spin-dependent gradient field is generated by two pairs of coils: an anti-Helmholtz pair along the $z$-direction, which creates a quadrupole field $B_\text{quad}\left(x\,\mathbf{e}_x+y\,\mathbf{e}_y-2z\,\mathbf{e}_z\right)$, and a Helmholtz pair along the $x$-direction, which creates a homogenous bias field~$B_x \mathbf{e}_x$ with $\mathbf{e}_i$ the unit vector in $i$-direction.
At the atom cloud ($x\approx0$), the total magnetic field is $B=\sqrt{(B_\text{quad}\,x+B_x)^2+B_\text{ofs}^2(y,z)}$ and the magnetic field gradient $B^\prime=\partial B/\partial x$ depends linearly on the bias field~$B_x$ for $B_x\ll B_\text{ofs}$.
A precise knowledge of the spin-dependent double well tilt is required for the current measurement method.
To this end, a calibration was performed using laser-assisted tunneling spectroscopy.
The setup comprises two beams interfered under an angle of $90^\circ$ with a frequency difference of $\delta\omega$ forming a running wave lattice oriented at $45^\circ$ to the physical lattice.
A single spin in the ($F=2$, $m_F=\pm 2$)-state is loaded in the ground state of a tilted double well potential with $\delta_0/h=4.91(2)\,\text{kHz}$ and additional magnetic tilt~$2\Delta$, which needs to be calibrated.
The running wave lattice modulates neighboring lattice sites relative to each other and will induce tunneling if $\hbar\,\delta\omega=2\Delta+\delta_0$.
A series of spectroscopy scans varying $\delta\omega$ are performed for various magnetic field values~$B_{x}$.
The resulting data is shown in \fig{fig:suppl:bfieldcalib}.
For large $B_{x}$ a deviation from the linear behavior is visible, which can be captured by fitting the magnetic field distribution of a quadrupole field
\begin{equation}
\Delta\propto\frac{B_{x}B_\text{quad}}{\sqrt{B_{x}^{2}+B_\text{ofs}^2}},
\label{eq:suppl:bfieldfit}
\end{equation}
where both $B_\text{quad}$ and $B_\text{ofs}$ are fit variables.

\newparagraph{Decay Calibration:}
\begin{figure}%
\includegraphics[width=\columnwidth]{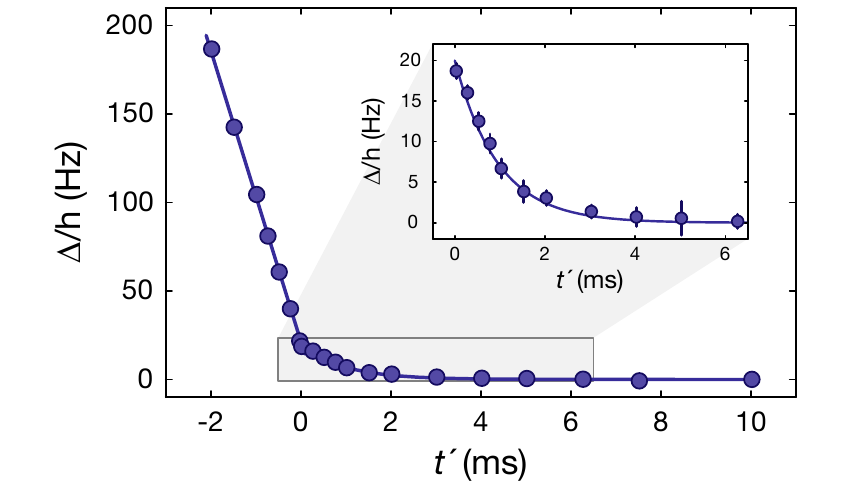}%
\caption{Calibration of the residual exponential decay of~$\Delta$ after the ramp stop $t^\prime=t-t_\text{s}$. The data points show the magnetic gradient determined by microwave spectroscopy on the  ($F=1$, $m_F=-1$)$\rightarrow{}$($F=2$, $m_F=-2$) transition when stopping the gradient ramp at $\Delta_\text{s}=0$. The linear ramp before and the exponential decay after the stop time~$t_\text{s}$ is clearly visible.
By fitting a linear~($t<0$) and an exponential function~($t>0$) to the data, the decay constant $\tau_\text{d}=1.05(5)\,\text{ms}$ and~$\Delta_\text{d}$, which depends on the double well site distance, can be determined.}%
\label{fig:suppl:bfielddecay}%
\end{figure}
During the current measurement sequence, the magnetic field gradient is ramped to a final value~$\Delta_\text{s}$ and abruptly stopped there. 
Experimentally, an instantaneous stop is not realizable but a small residual gradient remains which decays slowly during the current measurement.
The calibration of the decay time~$\tau_\text{d}$ as well as the residual gradient~$\Delta_\text{d}$ at $t_\text{s}$ is essential for the current measurement and is realized with microwave spectroscopy.
The sequence starts with a single spin in the ($F=1$, $m_F=-1$) state localized on the left site by a strong magnetic gradient. 
The gradient is reduced with the identical rate as for the current measurement and stopped at $\Delta_\text{s}=0$ but individual sites are decoupled by a large short lattice depth. 
Now, resonance frequency scans on the ($F=1$, $m_F=-1$)$\rightarrow$($F=2$, $m_F=-2$) microwave transition with a pulse duration of $44\,\upmu\text{s}$ are recorded at various times around~$t_\text{s}$; the gradient $\Delta(t^\prime)$ determined from the center frequencies is summarized in \fig{fig:suppl:bfielddecay}.
A clear exponential decay $\Delta_\text{d}\,e^{-t^\prime/\tau_\text{d}}$ can be fitted for $t^\prime>0$ with a decay constant~$\tau_\text{d}=1.05(5)\,\text{ms}$.
The amplitude~$\Delta_\text{d}$ can be calibrated by comparing the linear increase for $t^\prime<0$ with the gradient calibration (\fig{fig:suppl:bfieldcalib}) and leads to $\Delta_\text{d}/h=24.3(6)\,\text{Hz}$ for sites separated by $d_s$.
This corresponds to a tilt in the superlattice for the experimental parameters of $\Delta_\text{d}/h=19.8(5)\,\text{Hz}$ at $J_\text{ex}/h=342(2)\,\text{Hz}$ and $\Delta_\text{d}/h=19.4(5)\,\text{Hz}$ at $J_\text{ex}/h=467(3)\,\text{Hz}$.
The difference originates mainly in the slightly different distance between the sites depending on the double well parameters.

\newparagraph{Simultaneous band mapping of two spins:}
The simultaneous site-resolved detection of two spins per double well suffers from an additional reduction of the detected imbalance.
This reduction occurs during the merging of the left and right site of each double well in the presence of a spin-independent tilt  into a single site of the long lattice.
This process transfers atoms from the left site to the ground band and spins from the right site to the second excited band of the long lattice. 
The subsequent band mapping measures the band occupation and hence also the site occupations. 
However, spins in these bands undergo singlet-triplet oscillations after and also during the merging \cite{Anderlini:2007:Suppl,Trotzky:2007:Suppl}. 
A calibrated holdtime before the release is chosen such that correct imbalances are detected.
Nevertheless, the detected imbalance is reduced most likely due to dephasing during the merging ramps. 
The value of this reduction can be calibrated to rescale the measured imbalances with a constant factor to correctly determine the imbalance.
The calibration measurement compares the two spin site-resolved band mapping with a single spin band mapping, where shortly before the merging one of the spin components is removed by an adiabatic spin transfer and a subsequent resonant light pulse. 

\begin{figure}[t]%
\includegraphics[width=\columnwidth]{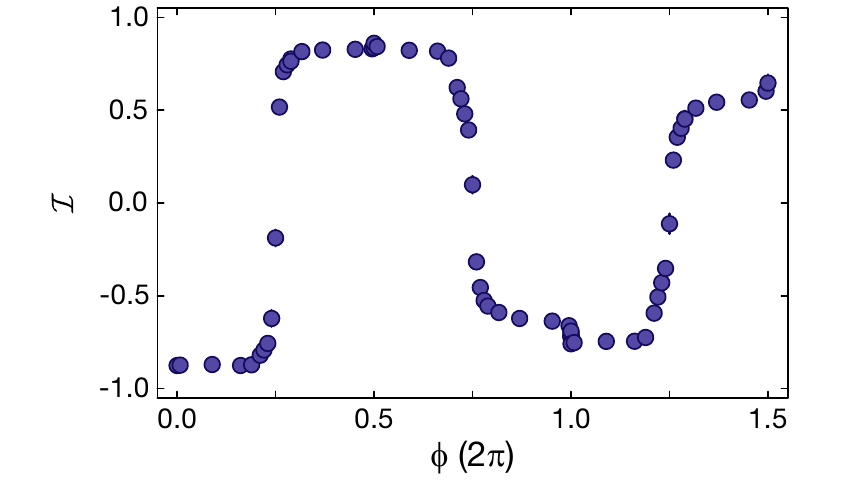}%
\caption{Static spin imbalance~$\mathcal{I}$ during the pump cycle. The data is shown with respect to the pump parameter~$\phi$ defined as the angle of the pump path in parameter space $(\delta J_\text{ex}/\delta J_\text{ex,max},\Delta/\Delta_\text{max})$. The error bars show the error of the mean of five repetitions each.}%
\label{fig:suppl:bandmapping}%
\end{figure}

\section{Multiple pump cycles}
\subsection{Band mapping data}
The in-situ data show a separation and opposite transport of the two spin components for multiple pump cycles.
In \fig{fig:suppl:bandmapping} the spin imbalance $\mathcal{I}_\text{s}$ during the pump cycle is depicted versus the pump parameter~$\phi$, which is defined as the angle of the pump path in parameter space $(\delta J_\text{ex}/\delta J_\text{ex,max},\Delta/\Delta_\text{max})$.
The imbalance starts with a negative value, the state is predominantly $\left|\uparrow\downarrow\right>$, and inverts during the first half pump cycle.
After switching the dimerization, the pump cycle continues by inverting the spin-imbalance each half pump cycle.
At~$\phi=2\pi$ a small step in~$\mathcal{I}$ is visible, which originates mainly from singly occupied sites created at the surface of the atom cloud during pumping.

\subsection{Initial state and pump efficiency}
For the in-situ measurement the calculated motion of a localized spin is shown, which takes into account the initial ground state occupation~$n_1^{\text{(i)}}$ and pump efficiency~$\beta_i$ per $i$-th half pump cycle.
The model is analogous to the one for the spin current, where the corrections are extracted from the spin-imbalance measured at each half pump cycle. 
The step height is then given by 
\begin{equation}
s_i=\beta_i\Bigl(2n_1^{\text{(i)}}\prod_{j=0}^{i-1}\beta_{j}-1\Bigr)
\label{eq:suppl:stepheight}
\end{equation}
with~$n_1^{\text{(i)}}=0.94$ and~$\beta_{0...4}=\{1,0.97,0.91,0.96,0.90\}$. In total, the displacement after the $i$-th half pump cycle is
\begin{equation}
x=\sum_{j=1} s_j.
\label{eq:suppl:titaldisplacement}
\end{equation}

\section{Pump scheme in a tight-binding model with magnetic field gradient}
In the tight-binding approximation the dynamics of non-interacting atoms in an optical superlattice potential in the presence of a field gradient is described by a generalized Harper model with a site-dependent Zeeman-like term
\begin{equation}
\begin{aligned}
\hat{H}_\text{s}=& \hat{H}_J+\hat{H}_\Delta\\
   =& - \sum_{m,\sigma} \frac{1}{2} \left(J+\delta J_m\right) (\hat{a}^\dagger_{m+1,\sigma}\hat{a}^{\phantom\dagger}_{m,\sigma} + \text{h.c.}) \\
		& + \sum_m m\,\Delta \left(\hat{a}^\dagger_{m,\uparrow} \hat{a}^{\phantom\dagger}_{m,\uparrow}-\hat{a}^\dagger_{m,\downarrow} \hat{a}^{\phantom\dagger}_{m,\downarrow}\right)
\end{aligned}
\label{eq:suppl:spin-harper}
\end{equation}
with $\delta J_m=(-1)^{m}\delta J$.
The pumping scheme is implemented with a cycle, in which $(\delta J,\ \Delta)\rightarrow(\delta J(\phi),\ \Delta(\phi))$ and where the pump parameter changes constantly in time $\phi=2\pi t/T$.
In the deep tight-binding regime the parameter space describes an ellipse $(\delta J(t),\ \Delta(t))=(\delta J \sin (2\pi t/T),\ \Delta \cos(2\pi t/T))$.
At $\Delta=0$, the spectrum has an energy gap $\Delta E=2\,\delta J$, and thus the adiabatic condition is met if $T \gg \hbar/\delta J$. 

The part of the Hamiltonian $\hat{H}_\Delta$ is odd while the part $\hat{H}_J$ is even under time-reversal symmetry and therefore \eq{eq:suppl:spin-harper} belongs to the class that satisfy the condition $\hat{H}[-t]=\hat{\Theta} \hat{H}[t] \hat{\Theta}^{-1}$, where $\hat{\Theta}$ is the time-reversal operator. 
Moreover, the Hamiltonian is time-reversal invariant at two points $t_1=\frac{T}{4}$ and $t_2=\frac{3T}{4}$, where $\hat{H}_J$ dominates. 
The existence of these two points plays a crucial role in the classification of the pump cycle.
In particular, pump cycles in which $\hat{H}[t_1]$ and $\hat{H}[t_2]$ have different time reversal polarization are topologically distinct from trivial cycles and define a $Z_2$ spin-pump. 
In a single double well at time $t=0$, $\hat{H}_\Delta$ dominates and locks the up (down) spins on the left (right) well, denoted by $\left|\uparrow, \downarrow \right>$.
This state evolves into the $\left|\downarrow, \uparrow\right>$ state after half a pump cycle at $t=\frac{T}{2}$, where the two spins have exchanged their positions. 
In contrast, at $t=\frac{T}{4}$ and $t=\frac{3T}{4}$, the term $\hat{H}_J$ dominates and the spins are delocalized over the double wells; then the system is dimerized. 

\section{Center of mass shift and time-reversal polarization}
Consider a Hamiltonian~\eq{eq:suppl:spin-harper} with lattice constant $d_s=d_l/2=1$ and periodic boundary conditions.
Then, in absence of spin-orbit type of interaction that means independent spin components without inter-spin interactions, the spin transport for a homogeneously populated band is characterized by the spin Chern number $C_\text{sc}=\nu_\uparrow-\nu_\downarrow$.
Since the spin components are decoupled even in the presence of a field gradient, the Chern numbers $\nu_\sigma$ can be evaluated using the Thouless-Kohmoto-Nightingale-Nijs expression \cite{Thouless:1982:Suppl}:
\begin{equation}
\nu_\sigma=\frac{1}{2\pi}\int_0^T \text{d}t \int_{-\pi}^{\pi} \text{d}k~\Omega_\sigma(t,k),
\end{equation}
where $\Omega_\sigma$ is the Berry curvature associated to the single-particle wavefunction
\begin{equation}
\Omega_\sigma(t,k)=\ii \left(\left< \partial_t u^\sigma\middle|\partial_k u^\sigma \right>-\text{h.c.}\right).
\end{equation}
The spin Chern number can be furthermore related for the non-interacting case to the $Z_2$ topological invariant $I=\text{mod}_2(C_\text{sc}/2)$ that distinguishes a nontrivial $Z_2$ pump from a trivial one and is related to the change in time reversal polarization.

As known from polarization theory \cite{Marzari:1997:Suppl}, the charge polarization is the center of mass of a localized Wannier state and is in turn related to Berry's phase of the corresponding Bloch functions.
In the same way, the polarization of a single spin component is given by:
\begin{equation}
P_\sigma=\frac{1}{2\pi}\int_{-\pi}^{\pi} \text{d}k~\mathcal{A}_\sigma(k),
\end{equation}
where $\mathcal{A}_\sigma(k)=\ii\sum \left< u_\sigma\middle|\partial_k u_\sigma\right>$ is the Berry connection. 
The change in polarization induced by changing the pump parameter $\phi$ by $2\pi$, or the time variable, corresponds to the Chern number~\cite{Marzari:1997:Suppl}
\begin{equation}
\nu_\sigma=\int_{0}^{2\pi} \text{d}\phi~\partial_\phi P_\sigma(\phi).
\label{eq:suppl:chern}
\end{equation}
The spin-density can be directly measured by in-situ absorption imaging for a single spin component. 
The change of spin-polarization $\Delta P_\sigma=P_\sigma (\phi_1)-P_\sigma (\phi_2)$ at two different times $t_1$ and $t_2$, coincides with the spatial shift of the Wannier function. 
Measuring the center of mass shift for a single spin component thus gives the Chern number of this component.

For time-reversal invariant systems, taking into account the role of Kramer's degeneracy, one can define a corresponding time-reversal polarization in terms of the difference of the individual spin polarizations $P_s=P_\uparrow-P_\downarrow$. Hence, the change in time-reversal polarization during a cycle gives the $Z_2$ topological invariant.
Furthermore, it is equal to the integration of the instantaneous spin-current~$\mathfrak{j}$ over the pump cycle as, $\int_0^{T} \text{d}t~\mathfrak{j}(t)$. 

\section{Spin pumping with interactions}
When in addition hardcore interactions between the spin components are assumed, spin pumping can be understood in a similar way as in the non-interacting case. 
For half filling a representation in terms of spin operators can be introduced in this limit:
\begin{equation}
\begin{aligned}
\hat{S}^+_m  &=  \hat{a}^\dagger_{m\uparrow}  \hat{a}^{\phantom\dagger}_{m\downarrow},\\
\hat{S}^-_m  &=  \hat{a}^\dagger_{m\downarrow}\hat{a}^{\phantom\dagger}_{m\uparrow},\\
\hat{S}^z_m  &=  \hat{a}^\dagger_{m\uparrow}\hat{a}^{\phantom\dagger}_{m\uparrow}-\hat{a}^\dagger_{m\downarrow}\hat{a}^{\phantom\dagger}_{m\downarrow}.
\end{aligned}
\end{equation}
Model \eq{eq:suppl:spin-harper} can thus be mapped to a model of an antiferromagnetic spin chain with two perturbations
\begin{equation}
\begin{aligned}
\hat{\mathcal{H}}_\text{eff}=&\hat{\mathcal{H}}_{xy}+\hat{\mathcal{H}}_\text{dim}+\hat{\mathcal{H}}_{\Delta}\\
						=&-\frac{J_\text{ex}}{4} \sum_m (\hat{S}^+_m \hat{S}^-_{m+1}+\text{h.c.})\\
						 &-\frac{\delta J_\text{ex}}{4}\sum_m (-1)^m (\hat{S}^+_m \hat{S}^-_{m+1}+\text{h.c.})\\
						 &+\Delta \sum_m m\,\hat{S}_m^z,
\end{aligned}
\label{eq:suppl:heisenberg}
\end{equation}
where the second term describes a staggered component of the exchange interaction, while the last one is a Zeeman-like term, which controls the on-site energies.

A cycle, in which $(\delta J_\text{ex},\ \Delta)$ are adiabatically varied defines a topological spin pump~\cite{Shindou:2005:Suppl}.
Such a spin pump transfers $S_z=\hbar$ per cycle, which can be still described by the $Z_2$ topological invariant.
Note that the Hamiltonian $\hat{\mathcal{H}}_{xy}$ in \eq{eq:suppl:heisenberg} corresponds to a cosine band $\epsilon (k)=-{J_\text{ex}}/2\,\cos (k)$.
The staggered exchange interaction $\delta J_\text{ex}$ opens a gap at $k=\pm \frac{\pi}{2}$ and for half filling, only the lowest subband is occupied.
Due to the $\pi$ periodicity in $k$-space, the double degenerate point $(k,~\delta J_\text{ex},~\Delta)=(\frac{\pi}{2},~0,~0)$ is identical to that at $(-\frac{\pi}{2}, 0,0)$ and becomes the source and sink for a vector field $\mathbf{B}_{+1}$ and $\mathbf{B}_{-1}$ defined in the $k$-$\phi$-parameter space.
If a pump path $\gamma$ encloses the origin $(\delta J_\text{ex},\ \Delta)=(0, 0)$, e.g.\ $(\delta J_\text{ex},\ \Delta)=(\delta J_\text{ex,max}\cos\phi,\ \Delta_\text{max}\sin \phi)$, where $\phi:0\rightarrow  2\pi$, the number of lattice sites that a spin is transported, is given by the flux $ \mathbf{B}_{+1}$  enclosed by the path
\begin{equation}
\oint_\gamma\int_{k=-\pi}^{\pi} \text{d}\mathbf{S }\cdot \mathbf{B}_{+1}=1.
\end{equation}
This corresponds to a quantized spin transport.
The total $S^z$ at one end of this system increases while that at the other end decreases by one during the entire cycle as long as the gap is maintained open and the point $(0,\ 0)$ is not outside the 2D closed surface.

Away from the hard-core constraints for the bosons, the effect of a finite interaction can be taken into account via a bosonization approach.
When applying Haldane's bosonization of interacting bosons~\cite{Haldane:1981:Suppl} to the Hamiltonian \eq{eq:suppl:spin-harper} and $\delta J_\text{ex}, \Delta=0$, the Hamiltonian of the bosons can be written as:
\begin{equation}
  \label{eq:suppl:2cll-ham}
  \hat{H}_0=\sum_\sigma \int \frac{dx}{2\pi} \left[ v_\sigma K_\sigma (\pi\,\Pi_\sigma)^2
      + \frac{v_\sigma}{K_\sigma} (\partial_x\Phi_\sigma)^2  \right],
\end{equation}
where the two canonical fields fulfill $[\Phi_\alpha(x),\Pi_\beta(x')]=\ii\,\delta_{\alpha\beta}\delta(x-x')$, $v_\sigma$ is the velocity of excitations, and $K_\sigma$ is the Tomonaga-Luttinger exponent. 
In the case of hard-core bosons, $v_\sigma = J_\text{ex} \sin (\pi \rho^{0}_\sigma)$ and $K_\sigma=1$, while $\rho^0$ is the boson density.

Introducing the fields $\theta_\alpha =\pi \int^x \Pi_\alpha$, the boson annihilation operators can be represented as~\cite{Haldane:1981:Suppl}:
\begin{align}
  a_{j\sigma}&=\psi_\sigma(x)\label{eq:suppl:boson-annihil}\\
	&=e^{\ii\theta_\sigma(x)} \sum_{m=0}^{+\infty} c^{\sigma}_m \cos(2m \Phi_\sigma(x) -2m \pi \rho^{(0)}_\sigma x),\nonumber
\end{align}
where $c^\sigma_m$ are non-universal coefficients.
For hardcore bosons at half filling, these coefficients have been found analytically~\cite{Ovchinnikov:2004:Suppl}.
From \eq{eq:suppl:boson-annihil}, the bosonized expression of the staggered hopping term can be deduced:
\begin{equation}
  \hat{H}_\text{hop.} \propto \delta J  \int \text{d}x \sin (2\Phi_c)\cos(2\Phi_s),
	\label{eq:suppl:boso-hopping}
\end{equation}
with only the most relevant term in the renormalization group sense and the charge and spin variables $\Phi_{\uparrow/\downarrow}=(\Phi_c \pm \Phi_s)$.
When $\Phi_c$ is pinned (e.g.\ at commensurate fillings) in the gapped spin phase also the field $\Phi_s$ is pinned $\langle \Phi_s\rangle \equiv \frac{\pi}{4}\left(1+\text{sign}(\delta J)\right)$. The excitations above the ground state are solitons and antisolitons, which 
are topological excitations of the field $\Phi_s$ that carry a spin $1/2$.

The time reversal-polarization is identified as $P_s=\text{mod}_2(\frac{2 \Phi_s}{\pi})$ and because under time reversal $\hat{\Theta}\Phi_s\hat{\Theta}^{-1}=-\Phi_s$, time reversal polarization is either $0$ or $1$. 
Thus, the topological classification of the spin pump remains also away from the hard-core bosons limit.
\bibliographystyle{bosons}

\begin{thebibliography}{34}%
\makeatletter
\providecommand \@ifxundefined [1]{%
 \@ifx{#1\undefined}
}%
\providecommand \@ifnum [1]{%
 \ifnum #1\expandafter \@firstoftwo
 \else \expandafter \@secondoftwo
 \fi
}%
\providecommand \@ifx [1]{%
 \ifx #1\expandafter \@firstoftwo
 \else \expandafter \@secondoftwo
 \fi
}%
\providecommand \natexlab [1]{#1}%
\providecommand \enquote  [1]{``#1''}%
\providecommand \bibnamefont  [1]{#1}%
\providecommand \bibfnamefont [1]{#1}%
\providecommand \citenamefont [1]{#1}%
\providecommand \href@noop [0]{\@secondoftwo}%
\providecommand \href [0]{\begingroup \@sanitize@url \@href}%
\providecommand \@href[1]{\@@startlink{#1}\@@href}%
\providecommand \@@href[1]{\endgroup#1\@@endlink}%
\providecommand \@sanitize@url [0]{\catcode `\\12\catcode `\$12\catcode
  `\&12\catcode `\#12\catcode `\^12\catcode `\_12\catcode `\%12\relax}%
\providecommand \@@startlink[1]{}%
\providecommand \@@endlink[0]{}%
\providecommand \url  [0]{\begingroup\@sanitize@url \@url }%
\providecommand \@url [1]{\endgroup\@href {#1}{\urlprefix }}%
\providecommand \urlprefix  [0]{URL }%
\providecommand \Eprint [0]{\href }%
\providecommand \doibase [0]{http://dx.doi.org/}%
\providecommand \selectlanguage [0]{\@gobble}%
\providecommand \bibinfo  [0]{\@secondoftwo}%
\providecommand \bibfield  [0]{\@secondoftwo}%
\providecommand \translation [1]{[#1]}%
\providecommand \BibitemOpen [0]{}%
\providecommand \bibitemStop [0]{}%
\providecommand \bibitemNoStop [0]{.\EOS\space}%
\providecommand \EOS [0]{\spacefactor3000\relax}%
\providecommand \BibitemShut  [1]{\csname bibitem#1\endcsname}%
\let\auto@bib@innerbib\@empty
%</preamble>
\bibitem [{\citenamefont {Klitzing et~al.}(1980)\citenamefont {Klitzing},
  \citenamefont {Dorda},\ and\ \citenamefont {Pepper}}]{Klitzing:1980}%
  \BibitemOpen
  \bibfield  {author} {\bibinfo {author} {\bibfnamefont {K.~v.}\ \bibnamefont
  {Klitzing}}, \bibinfo {author} {\bibfnamefont {G.}~\bibnamefont {Dorda}},\
  \bibnamefont {and}\ \bibinfo {author} {\bibfnamefont {M.}~\bibnamefont
  {Pepper}},\ }\href {\doibase 10.1103/PhysRevLett.45.494} {\bibfield
  {journal} {\bibinfo  {journal} {Phys. Rev. Lett.}\ }\textbf {\bibinfo
  {volume} {45}},\ \bibinfo {pages} {494}} (\bibinfo {year} {1980})\BibitemShut
  {NoStop}%
\bibitem [{\citenamefont {Tsui et~al.}(1982)\citenamefont {Tsui}, \citenamefont
  {Stormer},\ and\ \citenamefont {Gossard}}]{Tsui:1983}%
  \BibitemOpen
  \bibfield  {author} {\bibinfo {author} {\bibfnamefont {D.~C.}\ \bibnamefont
  {Tsui}}, \bibinfo {author} {\bibfnamefont {H.~L.}\ \bibnamefont {Stormer}},\
  \bibnamefont {and}\ \bibinfo {author} {\bibfnamefont {A.~C.}\ \bibnamefont
  {Gossard}},\ }\href {\doibase 10.1103/PhysRevLett.48.1559} {\bibfield
  {journal} {\bibinfo  {journal} {Phys. Rev. Lett.}\ }\textbf {\bibinfo
  {volume} {48}},\ \bibinfo {pages} {1559}} (\bibinfo {year}
  {1982})\BibitemShut {NoStop}%
\bibitem [{\citenamefont {Thouless et~al.}(1982)\citenamefont {Thouless},
  \citenamefont {Kohmoto}, \citenamefont {Nightingale},\ and\ \citenamefont
  {den Nijs}}]{Thouless:1982}%
  \BibitemOpen
  \bibfield  {author} {\bibinfo {author} {\bibfnamefont {D.~J.}\ \bibnamefont
  {Thouless}}, \bibinfo {author} {\bibfnamefont {M.}~\bibnamefont {Kohmoto}},
  \bibinfo {author} {\bibfnamefont {M.~P.}\ \bibnamefont {Nightingale}},\
  \bibnamefont {and}\ \bibinfo {author} {\bibfnamefont {M.}~\bibnamefont {den
  Nijs}},\ }\href {\doibase 10.1103/PhysRevLett.49.405} {\bibfield  {journal}
  {\bibinfo  {journal} {Phys. Rev. Lett.}\ }\textbf {\bibinfo {volume} {49}},\
  \bibinfo {pages} {405}} (\bibinfo {year} {1982})\BibitemShut {NoStop}%
\bibitem [{\citenamefont {Thouless}(1983)}]{Thouless:1983}%
  \BibitemOpen
  \bibfield  {author} {\bibinfo {author} {\bibfnamefont {D.~J.}\ \bibnamefont
  {Thouless}},\ }\href {\doibase 10.1103/PhysRevB.27.6083} {\bibfield
  {journal} {\bibinfo  {journal} {Phys. Rev. B}\ }\textbf {\bibinfo {volume}
  {27}},\ \bibinfo {pages} {6083}} (\bibinfo {year} {1983})\BibitemShut
  {NoStop}%
\bibitem [{\citenamefont {Niu\ and\ Thouless}(1984)\citenamefont {Niu}\ and\
  \citenamefont {Thouless}}]{Niu:1984}%
  \BibitemOpen
  \bibfield  {author} {\bibinfo {author} {\bibfnamefont {Q.}~\bibnamefont
  {Niu}}\ \bibnamefont {and}\ \bibinfo {author} {\bibfnamefont {D.~J.}\
  \bibnamefont {Thouless}},\ }\href
  {http://stacks.iop.org/0305-4470/17/i=12/a=016} {\bibfield  {journal}
  {\bibinfo  {journal} {J. Phys. A}\ }\textbf {\bibinfo {volume} {17}},\
  \bibinfo {pages} {2453}} (\bibinfo {year} {1984})\BibitemShut {NoStop}%
\bibitem [{\citenamefont {K{\"o}nig et~al.}(2007)\citenamefont {K{\"o}nig},
  \citenamefont {Wiedmann}, \citenamefont {Br{\"u}ne}, \citenamefont {Roth},
  \citenamefont {Buhmann}, \citenamefont {Molenkamp}, \citenamefont {Qi},\ and\
  \citenamefont {Zhang}}]{Koenig:2007}%
  \BibitemOpen
  \bibfield  {author} {\bibinfo {author} {\bibfnamefont {M.}~\bibnamefont
  {K{\"o}nig}}, \bibinfo {author} {\bibfnamefont {S.}~\bibnamefont {Wiedmann}},
  \bibinfo {author} {\bibfnamefont {C.}~\bibnamefont {Br{\"u}ne}}, \bibnamefont
  {et~al.},\ }\href {\doibase 10.1126/science.1148047} {\bibfield  {journal}
  {\bibinfo  {journal} {Science}\ }\textbf {\bibinfo {volume} {318}},\ \bibinfo
  {pages} {766}} (\bibinfo {year} {2007})\BibitemShut {NoStop}%
\bibitem [{\citenamefont {Kane\ and\ Mele}(2005)\citenamefont {Kane}\ and\
  \citenamefont {Mele}}]{Kane:2005}%
  \BibitemOpen
  \bibfield  {author} {\bibinfo {author} {\bibfnamefont {C.~L.}\ \bibnamefont
  {Kane}}\ \bibnamefont {and}\ \bibinfo {author} {\bibfnamefont {E.~J.}\
  \bibnamefont {Mele}},\ }\href {\doibase 10.1103/PhysRevLett.95.226801}
  {\bibfield  {journal} {\bibinfo  {journal} {Phys. Rev. Lett.}\ }\textbf
  {\bibinfo {volume} {95}},\ \bibinfo {pages} {226801}} (\bibinfo {year}
  {2005})\BibitemShut {NoStop}%
\bibitem [{\citenamefont {Bernevig\ and\ Zhang}(2006)\citenamefont {Bernevig}\
  and\ \citenamefont {Zhang}}]{Bernevig:2006}%
  \BibitemOpen
  \bibfield  {author} {\bibinfo {author} {\bibfnamefont {B.~A.}\ \bibnamefont
  {Bernevig}}\ \bibnamefont {and}\ \bibinfo {author} {\bibfnamefont {S.-C.}\
  \bibnamefont {Zhang}},\ }\href {\doibase 10.1103/PhysRevLett.96.106802}
  {\bibfield  {journal} {\bibinfo  {journal} {Phys. Rev. Lett.}\ }\textbf
  {\bibinfo {volume} {96}},\ \bibinfo {pages} {106802}} (\bibinfo {year}
  {2006})\BibitemShut {NoStop}%
\bibitem [{\citenamefont {Shindou}(2005)}]{Shindou:2005}%
  \BibitemOpen
  \bibfield  {author} {\bibinfo {author} {\bibfnamefont {R.}~\bibnamefont
  {Shindou}},\ }\href {\doibase 10.1143/JPSJ.74.1214} {\bibfield  {journal}
  {\bibinfo  {journal} {J. Phys. Soc. Jpn.}\ }\textbf {\bibinfo {volume}
  {74}},\ \bibinfo {pages} {1214}} (\bibinfo {year} {2005})\BibitemShut
  {NoStop}%
\bibitem [{\citenamefont {Fu\ and\ Kane}(2006)\citenamefont {Fu}\ and\
  \citenamefont {Kane}}]{Fu:2006}%
  \BibitemOpen
  \bibfield  {author} {\bibinfo {author} {\bibfnamefont {L.}~\bibnamefont
  {Fu}}\ \bibnamefont {and}\ \bibinfo {author} {\bibfnamefont {C.~L.}\
  \bibnamefont {Kane}},\ }\href {\doibase 10.1103/PhysRevB.74.195312}
  {\bibfield  {journal} {\bibinfo  {journal} {Phys. Rev. B}\ }\textbf {\bibinfo
  {volume} {74}},\ \bibinfo {pages} {195312}} (\bibinfo {year}
  {2006})\BibitemShut {NoStop}%
\bibitem [{\citenamefont {Zhou et~al.}(2014)\citenamefont {Zhou}, \citenamefont
  {Zhang}, \citenamefont {Sheng}, \citenamefont {Shen}, \citenamefont {Sheng},\
  and\ \citenamefont {Xing}}]{Zhou:2014}%
  \BibitemOpen
  \bibfield  {author} {\bibinfo {author} {\bibfnamefont {C.~Q.}\ \bibnamefont
  {Zhou}}, \bibinfo {author} {\bibfnamefont {Y.~F.}\ \bibnamefont {Zhang}},
  \bibinfo {author} {\bibfnamefont {L.}~\bibnamefont {Sheng}}, \bibnamefont
  {et~al.},\ }\href {\doibase 10.1103/PhysRevB.90.085133} {\bibfield  {journal}
  {\bibinfo  {journal} {Phys. Rev. B}\ }\textbf {\bibinfo {volume} {90}},\
  \bibinfo {pages} {085133}} (\bibinfo {year} {2014})\BibitemShut {NoStop}%
\bibitem [{\citenamefont {Prinz}(1998)}]{Prinz:1998}%
  \BibitemOpen
  \bibfield  {author} {\bibinfo {author} {\bibfnamefont {G.~A.}\ \bibnamefont
  {Prinz}},\ }\href {\doibase 10.1126/science.282.5394.1660} {\bibfield
  {journal} {\bibinfo  {journal} {Science}\ }\textbf {\bibinfo {volume}
  {282}},\ \bibinfo {pages} {1660}} (\bibinfo {year} {1998})\BibitemShut
  {NoStop}%
\bibitem [{\citenamefont {Murakami et~al.}(2003)\citenamefont {Murakami},
  \citenamefont {Nagaosa},\ and\ \citenamefont {Zhang}}]{Murakami:2003}%
  \BibitemOpen
  \bibfield  {author} {\bibinfo {author} {\bibfnamefont {S.}~\bibnamefont
  {Murakami}}, \bibinfo {author} {\bibfnamefont {N.}~\bibnamefont {Nagaosa}},\
  \bibnamefont {and}\ \bibinfo {author} {\bibfnamefont {S.-C.}\ \bibnamefont
  {Zhang}},\ }\href {\doibase 10.1126/science.1087128} {\bibfield  {journal}
  {\bibinfo  {journal} {Science}\ }\textbf {\bibinfo {volume} {301}},\ \bibinfo
  {pages} {1348}} (\bibinfo {year} {2003})\BibitemShut {NoStop}%
\bibitem [{\citenamefont {Kato et~al.}(2004)\citenamefont {Kato}, \citenamefont
  {Myers}, \citenamefont {Gossard},\ and\ \citenamefont
  {Awschalom}}]{Kato:2004}%
  \BibitemOpen
  \bibfield  {author} {\bibinfo {author} {\bibfnamefont {Y.~K.}\ \bibnamefont
  {Kato}}, \bibinfo {author} {\bibfnamefont {R.~C.}\ \bibnamefont {Myers}},
  \bibinfo {author} {\bibfnamefont {A.~C.}\ \bibnamefont {Gossard}},\
  \bibnamefont {and}\ \bibinfo {author} {\bibfnamefont {D.~D.}\ \bibnamefont
  {Awschalom}},\ }\href {\doibase 10.1126/science.1105514} {\bibfield
  {journal} {\bibinfo  {journal} {Science}\ }\textbf {\bibinfo {volume}
  {306}},\ \bibinfo {pages} {1910}} (\bibinfo {year} {2004})\BibitemShut
  {NoStop}%
\bibitem [{\citenamefont {Sharma\ and\ Chamon}(2001)\citenamefont {Sharma}\
  and\ \citenamefont {Chamon}}]{Sharma:2001}%
  \BibitemOpen
  \bibfield  {author} {\bibinfo {author} {\bibfnamefont {P.}~\bibnamefont
  {Sharma}}\ \bibnamefont {and}\ \bibinfo {author} {\bibfnamefont
  {C.}~\bibnamefont {Chamon}},\ }\href {\doibase 10.1103/PhysRevLett.87.096401}
  {\bibfield  {journal} {\bibinfo  {journal} {Phys. Rev. Lett.}\ }\textbf
  {\bibinfo {volume} {87}},\ \bibinfo {pages} {096401}} (\bibinfo {year}
  {2001})\BibitemShut {NoStop}%
\bibitem [{\citenamefont {Citro et~al.}(2003)\citenamefont {Citro},
  \citenamefont {Andrei},\ and\ \citenamefont {Niu}}]{Citro:2003}%
  \BibitemOpen
  \bibfield  {author} {\bibinfo {author} {\bibfnamefont {R.}~\bibnamefont
  {Citro}}, \bibinfo {author} {\bibfnamefont {N.}~\bibnamefont {Andrei}},\
  \bibnamefont {and}\ \bibinfo {author} {\bibfnamefont {Q.}~\bibnamefont
  {Niu}},\ }\href {\doibase 10.1103/PhysRevB.68.165312} {\bibfield  {journal}
  {\bibinfo  {journal} {Phys. Rev. B}\ }\textbf {\bibinfo {volume} {68}},\
  \bibinfo {pages} {165312}} (\bibinfo {year} {2003})\BibitemShut {NoStop}%
\bibitem [{\citenamefont {Citro et~al.}(2011)\citenamefont {Citro},
  \citenamefont {Romeo},\ and\ \citenamefont {Andrei}}]{Citro:2011}%
  \BibitemOpen
  \bibfield  {author} {\bibinfo {author} {\bibfnamefont {R.}~\bibnamefont
  {Citro}}, \bibinfo {author} {\bibfnamefont {F.}~\bibnamefont {Romeo}},\
  \bibnamefont {and}\ \bibinfo {author} {\bibfnamefont {N.}~\bibnamefont
  {Andrei}},\ }\href {\doibase 10.1103/PhysRevB.84.161301} {\bibfield
  {journal} {\bibinfo  {journal} {Phys. Rev. B}\ }\textbf {\bibinfo {volume}
  {84}},\ \bibinfo {pages} {161301}} (\bibinfo {year} {2011})\BibitemShut
  {NoStop}%
\bibitem [{\citenamefont {Watson et~al.}(2003)\citenamefont {Watson},
  \citenamefont {Potok}, \citenamefont {Marcus},\ and\ \citenamefont
  {Umansky}}]{Watson:2003}%
  \BibitemOpen
  \bibfield  {author} {\bibinfo {author} {\bibfnamefont {S.~K.}\ \bibnamefont
  {Watson}}, \bibinfo {author} {\bibfnamefont {R.~M.}\ \bibnamefont {Potok}},
  \bibinfo {author} {\bibfnamefont {C.~M.}\ \bibnamefont {Marcus}},\
  \bibnamefont {and}\ \bibinfo {author} {\bibfnamefont {V.}~\bibnamefont
  {Umansky}},\ }\href {\doibase 10.1103/PhysRevLett.91.258301} {\bibfield
  {journal} {\bibinfo  {journal} {Phys. Rev. Lett.}\ }\textbf {\bibinfo
  {volume} {91}},\ \bibinfo {pages} {258301}} (\bibinfo {year}
  {2003})\BibitemShut {NoStop}%
\bibitem [{\citenamefont {Sandweg et~al.}(2011)\citenamefont {Sandweg},
  \citenamefont {Kajiwara}, \citenamefont {Chumak}, \citenamefont {Serga},
  \citenamefont {Vasyuchka}, \citenamefont {Jungfleisch}, \citenamefont
  {Saitoh},\ and\ \citenamefont {Hillebrands}}]{Sandweg:2011}%
  \BibitemOpen
  \bibfield  {author} {\bibinfo {author} {\bibfnamefont {C.~W.}\ \bibnamefont
  {Sandweg}}, \bibinfo {author} {\bibfnamefont {Y.}~\bibnamefont {Kajiwara}},
  \bibinfo {author} {\bibfnamefont {A.~V.}\ \bibnamefont {Chumak}},
  \bibnamefont {et~al.},\ }\href {\doibase 10.1103/PhysRevLett.106.216601}
  {\bibfield  {journal} {\bibinfo  {journal} {Phys. Rev. Lett.}\ }\textbf
  {\bibinfo {volume} {106}},\ \bibinfo {pages} {216601}} (\bibinfo {year}
  {2011})\BibitemShut {NoStop}%
\bibitem [{\citenamefont {Lohse et~al.}(2016)\citenamefont {Lohse},
  \citenamefont {Schweizer}, \citenamefont {Zilberberg}, \citenamefont
  {Aidelsburger},\ and\ \citenamefont {Bloch}}]{Lohse:2016}%
  \BibitemOpen
  \bibfield  {author} {\bibinfo {author} {\bibfnamefont {M.}~\bibnamefont
  {Lohse}}, \bibinfo {author} {\bibfnamefont {C.}~\bibnamefont {Schweizer}},
  \bibinfo {author} {\bibfnamefont {O.}~\bibnamefont {Zilberberg}}, \bibinfo
  {author} {\bibfnamefont {M.}~\bibnamefont {Aidelsburger}},\ \bibnamefont
  {and}\ \bibinfo {author} {\bibfnamefont {I.}~\bibnamefont {Bloch}},\ }\href
  {http://dx.doi.org/10.1038/nphys3584} {\bibfield  {journal} {\bibinfo
  {journal} {Nat. Phys.}\ }\textbf {\bibinfo {volume} {12}},\ \bibinfo {pages}
  {350}} (\bibinfo {year} {2016})\BibitemShut {NoStop}%
\bibitem [{\citenamefont {Lu et~al.}(2016)\citenamefont {Lu}, \citenamefont
  {Schemmer}, \citenamefont {Aycock}, \citenamefont {Genkina}, \citenamefont
  {Sugawa},\ and\ \citenamefont {Spielman}}]{Lu:2016}%
  \BibitemOpen
  \bibfield  {author} {\bibinfo {author} {\bibfnamefont {H.-I.}\ \bibnamefont
  {Lu}}, \bibinfo {author} {\bibfnamefont {M.}~\bibnamefont {Schemmer}},
  \bibinfo {author} {\bibfnamefont {L.~M.}\ \bibnamefont {Aycock}},
  \bibnamefont {et~al.},\ }\href {\doibase 10.1103/PhysRevLett.116.200402}
  {\bibfield  {journal} {\bibinfo  {journal} {Phys. Rev. Lett.}\ }\textbf
  {\bibinfo {volume} {116}},\ \bibinfo {pages} {200402}} (\bibinfo {year}
  {2016})\BibitemShut {NoStop}%
\bibitem [{\citenamefont {Nakajima et~al.}(2016)\citenamefont {Nakajima},
  \citenamefont {Tomita}, \citenamefont {Taie}, \citenamefont {Ichinose},
  \citenamefont {Ozawa}, \citenamefont {Wang}, \citenamefont {Troyer},\ and\
  \citenamefont {Takahashi}}]{Nakajima:2016}%
  \BibitemOpen
  \bibfield  {author} {\bibinfo {author} {\bibfnamefont {S.}~\bibnamefont
  {Nakajima}}, \bibinfo {author} {\bibfnamefont {T.}~\bibnamefont {Tomita}},
  \bibinfo {author} {\bibfnamefont {S.}~\bibnamefont {Taie}}, \bibnamefont
  {et~al.},\ }\href {http://dx.doi.org/10.1038/nphys3622} {\bibfield  {journal}
  {\bibinfo  {journal} {Nat Phys}\ }\textbf {\bibinfo {volume} {12}},\ \bibinfo
  {pages} {296}} (\bibinfo {year} {2016})\BibitemShut {NoStop}%
\bibitem [{\citenamefont {Rice\ and\ Mele}(1982)\citenamefont {Rice}\ and\
  \citenamefont {Mele}}]{Rice:1982}%
  \BibitemOpen
  \bibfield  {author} {\bibinfo {author} {\bibfnamefont {M.~J.}\ \bibnamefont
  {Rice}}\ \bibnamefont {and}\ \bibinfo {author} {\bibfnamefont {E.~J.}\
  \bibnamefont {Mele}},\ }\href {\doibase 10.1103/PhysRevLett.49.1455}
  {\bibfield  {journal} {\bibinfo  {journal} {Phys. Rev. Lett.}\ }\textbf
  {\bibinfo {volume} {49}},\ \bibinfo {pages} {1455}} (\bibinfo {year}
  {1982})\BibitemShut {NoStop}%
\bibitem [{\citenamefont {Lee et~al.}(2007)\citenamefont {Lee}, \citenamefont
  {Anderlini}, \citenamefont {Brown}, \citenamefont {Sebby-Strabley},
  \citenamefont {Phillips},\ and\ \citenamefont {Porto}}]{Lee:2007}%
  \BibitemOpen
  \bibfield  {author} {\bibinfo {author} {\bibfnamefont {P.~J.}\ \bibnamefont
  {Lee}}, \bibinfo {author} {\bibfnamefont {M.}~\bibnamefont {Anderlini}},
  \bibinfo {author} {\bibfnamefont {B.~L.}\ \bibnamefont {Brown}}, \bibnamefont
  {et~al.},\ }\href {\doibase 10.1103/PhysRevLett.99.020402} {\bibfield
  {journal} {\bibinfo  {journal} {Phys. Rev. Lett.}\ }\textbf {\bibinfo
  {volume} {99}},\ \bibinfo {pages} {020402}} (\bibinfo {year}
  {2007})\BibitemShut {NoStop}%
\bibitem [{\citenamefont {Dai et~al.}(2016)\citenamefont {Dai}, \citenamefont
  {Yang}, \citenamefont {Reingruber}, \citenamefont {Xu}, \citenamefont
  {Jiang}, \citenamefont {Chen}, \citenamefont {Yuan},\ and\ \citenamefont
  {Pan}}]{Dai:2015}%
  \BibitemOpen
  \bibfield  {author} {\bibinfo {author} {\bibfnamefont {H.-N.}\ \bibnamefont
  {Dai}}, \bibinfo {author} {\bibfnamefont {B.}~\bibnamefont {Yang}}, \bibinfo
  {author} {\bibfnamefont {A.}~\bibnamefont {Reingruber}}, \bibnamefont
  {et~al.},\ }\href {http://dx.doi.org/10.1038/nphys3705} {\bibfield  {journal}
  {\bibinfo  {journal} {Nat. Phys.}\ }\textbf {\bibinfo {volume} {12}},\
  \bibinfo {pages} {783}} (\bibinfo {year} {2016}),\ \bibinfo {note}
  {article}\BibitemShut {NoStop}%
\bibitem [{Sup()}]{Supplementary}%
  \BibitemOpen
  \href@noop {} {}\bibinfo {note} {See Supplemental Material}\BibitemShut
  {NoStop}%
\bibitem [{\citenamefont {Widera et~al.}(2005)\citenamefont {Widera},
  \citenamefont {Gerbier}, \citenamefont {F\"olling}, \citenamefont {Gericke},
  \citenamefont {Mandel},\ and\ \citenamefont {Bloch}}]{Widera:2005}%
  \BibitemOpen
  \bibfield  {author} {\bibinfo {author} {\bibfnamefont {A.}~\bibnamefont
  {Widera}}, \bibinfo {author} {\bibfnamefont {F.}~\bibnamefont {Gerbier}},
  \bibinfo {author} {\bibfnamefont {S.}~\bibnamefont {F\"olling}}, \bibnamefont
  {et~al.},\ }\href {\doibase 10.1103/PhysRevLett.95.190405} {\bibfield
  {journal} {\bibinfo  {journal} {Phys. Rev. Lett.}\ }\textbf {\bibinfo
  {volume} {95}},\ \bibinfo {pages} {190405}} (\bibinfo {year}
  {2005})\BibitemShut {NoStop}%
\bibitem [{\citenamefont {Xiao et~al.}(2010)\citenamefont {Xiao}, \citenamefont
  {Chang},\ and\ \citenamefont {Niu}}]{Xiao:2010}%
  \BibitemOpen
  \bibfield  {author} {\bibinfo {author} {\bibfnamefont {D.}~\bibnamefont
  {Xiao}}, \bibinfo {author} {\bibfnamefont {M.-C.}\ \bibnamefont {Chang}},\
  \bibnamefont {and}\ \bibinfo {author} {\bibfnamefont {Q.}~\bibnamefont
  {Niu}},\ }\href {\doibase 10.1103/RevModPhys.82.1959} {\bibfield  {journal}
  {\bibinfo  {journal} {Rev. Mod. Phys.}\ }\textbf {\bibinfo {volume} {82}},\
  \bibinfo {pages} {1959}} (\bibinfo {year} {2010})\BibitemShut {NoStop}%
\bibitem [{\citenamefont {Trotzky et~al.}(2008)\citenamefont {Trotzky},
  \citenamefont {Cheinet}, \citenamefont {F{\"o}lling}, \citenamefont {Feld},
  \citenamefont {Schnorrberger}, \citenamefont {Rey}, \citenamefont
  {Polkovnikov}, \citenamefont {Demler}, \citenamefont {Lukin},\ and\
  \citenamefont {Bloch}}]{Trotzky:2007}%
  \BibitemOpen
  \bibfield  {author} {\bibinfo {author} {\bibfnamefont {S.}~\bibnamefont
  {Trotzky}}, \bibinfo {author} {\bibfnamefont {P.}~\bibnamefont {Cheinet}},
  \bibinfo {author} {\bibfnamefont {S.}~\bibnamefont {F{\"o}lling}},
  \bibnamefont {et~al.},\ }\href {\doibase 10.1126/science.1150841} {\bibfield
  {journal} {\bibinfo  {journal} {Science}\ }\textbf {\bibinfo {volume}
  {319}},\ \bibinfo {pages} {295}} (\bibinfo {year} {2008})\BibitemShut
  {NoStop}%
\bibitem [{\citenamefont {Scarola\ and\ Das~Sarma}(2005)\citenamefont
  {Scarola}\ and\ \citenamefont {Das~Sarma}}]{Scarola:2005}%
  \BibitemOpen
  \bibfield  {author} {\bibinfo {author} {\bibfnamefont {V.~W.}\ \bibnamefont
  {Scarola}}\ \bibnamefont {and}\ \bibinfo {author} {\bibfnamefont
  {S.}~\bibnamefont {Das~Sarma}},\ }\href {\doibase
  10.1103/PhysRevLett.95.033003} {\bibfield  {journal} {\bibinfo  {journal}
  {Phys. Rev. Lett.}\ }\textbf {\bibinfo {volume} {95}},\ \bibinfo {pages}
  {033003}} (\bibinfo {year} {2005})\BibitemShut {NoStop}%
\bibitem [{\citenamefont {Hatsugai\ and\ Fukui}(2016)\citenamefont {Hatsugai}\
  and\ \citenamefont {Fukui}}]{Hatsugai:2016}%
  \BibitemOpen
  \bibfield  {author} {\bibinfo {author} {\bibfnamefont {Y.}~\bibnamefont
  {Hatsugai}}\ \bibnamefont {and}\ \bibinfo {author} {\bibfnamefont
  {T.}~\bibnamefont {Fukui}},\ }\href {\doibase 10.1103/PhysRevB.94.041102}
  {\bibfield  {journal} {\bibinfo  {journal} {Phys. Rev. B}\ }\textbf {\bibinfo
  {volume} {94}},\ \bibinfo {pages} {041102}} (\bibinfo {year}
  {2016})\BibitemShut {NoStop}%
\bibitem [{\citenamefont {Zhou}(2004)}]{Zhou:2004}%
  \BibitemOpen
  \bibfield  {author} {\bibinfo {author} {\bibfnamefont {F.}~\bibnamefont
  {Zhou}},\ }\href {\doibase 10.1103/PhysRevB.70.125321} {\bibfield  {journal}
  {\bibinfo  {journal} {Phys. Rev. B}\ }\textbf {\bibinfo {volume} {70}},\
  \bibinfo {pages} {125321}} (\bibinfo {year} {2004})\BibitemShut {NoStop}%
\bibitem [{\citenamefont {Bustos-Mar\'un et~al.}(2013)\citenamefont
  {Bustos-Mar\'un}, \citenamefont {Refael},\ and\ \citenamefont {von
  Oppen}}]{Bustos-Marun:2013}%
  \BibitemOpen
  \bibfield  {author} {\bibinfo {author} {\bibfnamefont {R.}~\bibnamefont
  {Bustos-Mar\'un}}, \bibinfo {author} {\bibfnamefont {G.}~\bibnamefont
  {Refael}},\ \bibnamefont {and}\ \bibinfo {author} {\bibfnamefont
  {F.}~\bibnamefont {von Oppen}},\ }\href {\doibase
  10.1103/PhysRevLett.111.060802} {\bibfield  {journal} {\bibinfo  {journal}
  {Phys. Rev. Lett.}\ }\textbf {\bibinfo {volume} {111}},\ \bibinfo {pages}
  {060802}} (\bibinfo {year} {2013})\BibitemShut {NoStop}%
\bibitem [{\citenamefont {Meidan et~al.}(2011)\citenamefont {Meidan},
  \citenamefont {Micklitz},\ and\ \citenamefont {Brouwer}}]{Meidan:2011}%
  \BibitemOpen
  \bibfield  {author} {\bibinfo {author} {\bibfnamefont {D.}~\bibnamefont
  {Meidan}}, \bibinfo {author} {\bibfnamefont {T.}~\bibnamefont {Micklitz}},\
  \bibnamefont {and}\ \bibinfo {author} {\bibfnamefont {P.~W.}\ \bibnamefont
  {Brouwer}},\ }\href {\doibase 10.1103/PhysRevB.84.075325} {\bibfield
  {journal} {\bibinfo  {journal} {Phys. Rev. B}\ }\textbf {\bibinfo {volume}
  {84}},\ \bibinfo {pages} {075325}} (\bibinfo {year} {2011})\BibitemShut
  {NoStop}%
\end{thebibliography}

\begin{thebibliography}{10}%
\makeatletter
\providecommand \@ifxundefined [1]{%
 \@ifx{#1\undefined}
}%
\providecommand \@ifnum [1]{%
 \ifnum #1\expandafter \@firstoftwo
 \else \expandafter \@secondoftwo
 \fi
}%
\providecommand \@ifx [1]{%
 \ifx #1\expandafter \@firstoftwo
 \else \expandafter \@secondoftwo
 \fi
}%
\providecommand \natexlab [1]{#1}%
\providecommand \enquote  [1]{``#1''}%
\providecommand \bibnamefont  [1]{#1}%
\providecommand \bibfnamefont [1]{#1}%
\providecommand \citenamefont [1]{#1}%
\providecommand \href@noop [0]{\@secondoftwo}%
\providecommand \href [0]{\begingroup \@sanitize@url \@href}%
\providecommand \@href[1]{\@@startlink{#1}\@@href}%
\providecommand \@@href[1]{\endgroup#1\@@endlink}%
\providecommand \@sanitize@url [0]{\catcode `\\12\catcode `\$12\catcode
  `\&12\catcode `\#12\catcode `\^12\catcode `\_12\catcode `\%12\relax}%
\providecommand \@@startlink[1]{}%
\providecommand \@@endlink[0]{}%
\providecommand \url  [0]{\begingroup\@sanitize@url \@url }%
\providecommand \@url [1]{\endgroup\@href {#1}{\urlprefix }}%
\providecommand \urlprefix  [0]{URL }%
\providecommand \Eprint [0]{\href }%
\providecommand \doibase [0]{http://dx.doi.org/}%
\providecommand \selectlanguage [0]{\@gobble}%
\providecommand \bibinfo  [0]{\@secondoftwo}%
\providecommand \bibfield  [0]{\@secondoftwo}%
\providecommand \translation [1]{[#1]}%
\providecommand \BibitemOpen [0]{}%
\providecommand \bibitemStop [0]{}%
\providecommand \bibitemNoStop [0]{.\EOS\space}%
\providecommand \EOS [0]{\spacefactor3000\relax}%
\providecommand \BibitemShut  [1]{\csname bibitem#1\endcsname}%
\let\auto@bib@innerbib\@empty
%</preamble>
\bibitem [{\citenamefont {Scarola\ and\ Das~Sarma}(2005)\citenamefont
  {Scarola}\ and\ \citenamefont {Das~Sarma}}]{Scarola:2005:Suppl}%
  \BibitemOpen
  \bibfield  {author} {\bibinfo {author} {\bibfnamefont {V.~W.}\ \bibnamefont
  {Scarola}}\ \bibnamefont {and}\ \bibinfo {author} {\bibfnamefont
  {S.}~\bibnamefont {Das~Sarma}},\ }\href {\doibase
  10.1103/PhysRevLett.95.033003} {\bibfield  {journal} {\bibinfo  {journal}
  {Phys. Rev. Lett.}\ }\textbf {\bibinfo {volume} {95}},\ \bibinfo {pages}
  {033003}} (\bibinfo {year} {2005})\BibitemShut {NoStop}%
\bibitem [{\citenamefont {Trotzky et~al.}(2008)\citenamefont {Trotzky},
  \citenamefont {Cheinet}, \citenamefont {F{\"o}lling}, \citenamefont {Feld},
  \citenamefont {Schnorrberger}, \citenamefont {Rey}, \citenamefont
  {Polkovnikov}, \citenamefont {Demler}, \citenamefont {Lukin},\ and\
  \citenamefont {Bloch}}]{Trotzky:2007:Suppl}%
  \BibitemOpen
  \bibfield  {author} {\bibinfo {author} {\bibfnamefont {S.}~\bibnamefont
  {Trotzky}}, \bibinfo {author} {\bibfnamefont {P.}~\bibnamefont {Cheinet}},
  \bibinfo {author} {\bibfnamefont {S.}~\bibnamefont {F{\"o}lling}},
  \bibnamefont {et~al.},\ }\href {\doibase 10.1126/science.1150841} {\bibfield
  {journal} {\bibinfo  {journal} {Science}\ }\textbf {\bibinfo {volume}
  {319}},\ \bibinfo {pages} {295}} (\bibinfo {year} {2008})\BibitemShut
  {NoStop}%
\bibitem [{\citenamefont {Xiao et~al.}(2010)\citenamefont {Xiao}, \citenamefont
  {Chang},\ and\ \citenamefont {Niu}}]{Xiao:2010:Suppl}%
  \BibitemOpen
  \bibfield  {author} {\bibinfo {author} {\bibfnamefont {D.}~\bibnamefont
  {Xiao}}, \bibinfo {author} {\bibfnamefont {M.-C.}\ \bibnamefont {Chang}},\
  \bibnamefont {and}\ \bibinfo {author} {\bibfnamefont {Q.}~\bibnamefont
  {Niu}},\ }\href {\doibase 10.1103/RevModPhys.82.1959} {\bibfield  {journal}
  {\bibinfo  {journal} {Rev. Mod. Phys.}\ }\textbf {\bibinfo {volume} {82}},\
  \bibinfo {pages} {1959}} (\bibinfo {year} {2010})\BibitemShut {NoStop}%
\bibitem [{\citenamefont {Widera et~al.}(2005)\citenamefont {Widera},
  \citenamefont {Gerbier}, \citenamefont {F\"olling}, \citenamefont {Gericke},
  \citenamefont {Mandel},\ and\ \citenamefont {Bloch}}]{Widera:2005:Suppl}%
  \BibitemOpen
  \bibfield  {author} {\bibinfo {author} {\bibfnamefont {A.}~\bibnamefont
  {Widera}}, \bibinfo {author} {\bibfnamefont {F.}~\bibnamefont {Gerbier}},
  \bibinfo {author} {\bibfnamefont {S.}~\bibnamefont {F\"olling}}, \bibnamefont
  {et~al.},\ }\href {\doibase 10.1103/PhysRevLett.95.190405} {\bibfield
  {journal} {\bibinfo  {journal} {Phys. Rev. Lett.}\ }\textbf {\bibinfo
  {volume} {95}},\ \bibinfo {pages} {190405}} (\bibinfo {year}
  {2005})\BibitemShut {NoStop}%
\bibitem [{\citenamefont {Anderlini et~al.}(2007)\citenamefont {Anderlini},
  \citenamefont {Lee}, \citenamefont {Brown}, \citenamefont {Sebby-Strabley},
  \citenamefont {Phillips},\ and\ \citenamefont
  {Porto}}]{Anderlini:2007:Suppl}%
  \BibitemOpen
  \bibfield  {author} {\bibinfo {author} {\bibfnamefont {M.}~\bibnamefont
  {Anderlini}}, \bibinfo {author} {\bibfnamefont {P.~J.}\ \bibnamefont {Lee}},
  \bibinfo {author} {\bibfnamefont {B.~L.}\ \bibnamefont {Brown}}, \bibnamefont
  {et~al.},\ }\href {\doibase 10.1038/nature06011} {\bibfield  {journal}
  {\bibinfo  {journal} {Nature}\ }\textbf {\bibinfo {volume} {448}},\ \bibinfo
  {pages} {452}} (\bibinfo {year} {2007})\BibitemShut {NoStop}%
\bibitem [{\citenamefont {Thouless et~al.}(1982)\citenamefont {Thouless},
  \citenamefont {Kohmoto}, \citenamefont {Nightingale},\ and\ \citenamefont
  {den Nijs}}]{Thouless:1982:Suppl}%
  \BibitemOpen
  \bibfield  {author} {\bibinfo {author} {\bibfnamefont {D.~J.}\ \bibnamefont
  {Thouless}}, \bibinfo {author} {\bibfnamefont {M.}~\bibnamefont {Kohmoto}},
  \bibinfo {author} {\bibfnamefont {M.~P.}\ \bibnamefont {Nightingale}},\
  \bibnamefont {and}\ \bibinfo {author} {\bibfnamefont {M.}~\bibnamefont {den
  Nijs}},\ }\href {\doibase 10.1103/PhysRevLett.49.405} {\bibfield  {journal}
  {\bibinfo  {journal} {Phys. Rev. Lett.}\ }\textbf {\bibinfo {volume} {49}},\
  \bibinfo {pages} {405}} (\bibinfo {year} {1982})\BibitemShut {NoStop}%
\bibitem [{\citenamefont {Marzari\ and\ Vanderbilt}(1997)\citenamefont
  {Marzari}\ and\ \citenamefont {Vanderbilt}}]{Marzari:1997:Suppl}%
  \BibitemOpen
  \bibfield  {author} {\bibinfo {author} {\bibfnamefont {N.}~\bibnamefont
  {Marzari}}\ \bibnamefont {and}\ \bibinfo {author} {\bibfnamefont
  {D.}~\bibnamefont {Vanderbilt}},\ }\href {\doibase 10.1103/PhysRevB.56.12847}
  {\bibfield  {journal} {\bibinfo  {journal} {Phys. Rev. B}\ }\textbf {\bibinfo
  {volume} {56}},\ \bibinfo {pages} {12847}} (\bibinfo {year}
  {1997})\BibitemShut {NoStop}%
\bibitem [{\citenamefont {Shindou}(2005)}]{Shindou:2005:Suppl}%
  \BibitemOpen
  \bibfield  {author} {\bibinfo {author} {\bibfnamefont {R.}~\bibnamefont
  {Shindou}},\ }\href {\doibase 10.1143/JPSJ.74.1214} {\bibfield  {journal}
  {\bibinfo  {journal} {J. Phys. Soc. Jpn.}\ }\textbf {\bibinfo {volume}
  {74}},\ \bibinfo {pages} {1214}} (\bibinfo {year} {2005})\BibitemShut
  {NoStop}%
\bibitem [{\citenamefont {Haldane}(1981)}]{Haldane:1981:Suppl}%
  \BibitemOpen
  \bibfield  {author} {\bibinfo {author} {\bibfnamefont {F.~D.~M.}\
  \bibnamefont {Haldane}},\ }\href {\doibase 10.1103/PhysRevLett.47.1840}
  {\bibfield  {journal} {\bibinfo  {journal} {Phys. Rev. Lett.}\ }\textbf
  {\bibinfo {volume} {47}},\ \bibinfo {pages} {1840}} (\bibinfo {year}
  {1981})\BibitemShut {NoStop}%
\bibitem [{\citenamefont {Ovchinnikov}(2004)}]{Ovchinnikov:2004:Suppl}%
  \BibitemOpen
  \bibfield  {author} {\bibinfo {author} {\bibfnamefont {A.~A.}\ \bibnamefont
  {Ovchinnikov}},\ }\href {\doibase 10.1088/0953-8984/16/18/016} {\bibfield
  {journal} {\bibinfo  {journal} {J. Phys. Condens. Matter}\ }\textbf {\bibinfo
  {volume} {16}},\ \bibinfo {pages} {3147}} (\bibinfo {year}
  {2004})\BibitemShut {NoStop}%
\end{thebibliography}

\end{document}